\documentclass[letterpaper,titlepage,11pt]{article}
\usepackage{hyperref}
\usepackage{amssymb,amsmath,amsfonts}
\usepackage{epsfig}
\usepackage{graphicx}
\usepackage{float}
\usepackage{subfig}
\usepackage{wrapfig}
\usepackage[english]{babel}
\usepackage[T1]{fontenc}
\usepackage[applemac]{inputenc}

\def\clock{{\count0=\time
           \divide\count0 60
           \ifnum\count0<10 0\fi\the\count0
           \multiply\count0 -60 \advance\count0 \time
           :\ifnum\count0<10 0\fi \the\count0
         }}
\newcommand{\timestamp}{{\small\vbox{\hbox{\tt\jobname.tex}
\hbox{\the\day/\the\month/\the\year, \clock}}}}

\newtheorem{definition}{Definition}[section]
\newtheorem{proposition}[definition]{Proposition}
\newtheorem{theorem}[definition]{Theorem}
\newtheorem{lemma}[definition]{Lemma}

\numberwithin{equation}{section}

\begin{document}

\begin{titlepage}

\vskip 2cm

\centerline{\Huge \bf  Maximal Analytic Extension and Hidden}
\vskip 0.4cm
\centerline{\Huge \bf  Symmetries of the Dipole Black Ring}
\vskip 1.4cm
\centerline{\bf  Jay Armas}
\vskip 0.3cm
\centerline{\sl The Niels Bohr Institute}
\centerline{\sl Blegdamsvej 17, DK-2100 Copenhagen \O, Denmark}
\vskip 0.1cm
\centerline{\sl and}
\vskip 0.1cm
\centerline{\sl Fabrikken}
\centerline{\sl Fabriksområdet 99, DK-1440 Copenhagen K, Christiania}
\vskip 0.1cm
\centerline{\sl and}
\vskip 0.1cm
\centerline{\sl Tata Institute of Fundamental Research}
\centerline{\sl Homi Bhabha Road, Mumbai 400005, India}
\vskip 0.4cm
\centerline{\small\tt jay@nbi.dk}

\vskip 2cm

\centerline{\bf Abstract} \vskip 0.2cm \noindent We construct analytic extensions across the Killing horizons of non-extremal and extremal dipole black rings in Einstein-Maxwell's theory using different methods. We show that these extensions are non-globally hyperbolic, have multiple asymptotically flat regions and in the non-extremal case, are also maximal and timelike complete. Moreover, we find that in both cases the causal structure of the maximally extended space-time resembles that of the 4-dimensional Reissner-Nordstr\"{o}m black hole. Furthermore, motivated by the physical interpretation of one of these extensions, we find a separable solution to the Hamilton-Jacobi equation corresponding to zero energy null geodesics and relate it to the existence of a conformal Killing tensor and a conformal Killing-Yano tensor in a specific dimensionally reduced space-time.

\vskip 3.5cm

\begin{flushright}
\emph{"Bevar Christiania"} 
\end{flushright}

\end{titlepage}

\small
\tableofcontents
\normalsize
\setcounter{page}{1}


\section{A short introduction} \label{intro}
In the past few years the interest in higher dimensional gravity has increased considerably as it was realized that, taking the dimension of the space-time $d$ as a parameter, the space of possible solutions becomes more complex, intricate and richer as $d$ takes higher values. In particular, in 5-dimensional vacuum gravity many exact analytic solutions were found describing different black hole configurations \cite{Emparan:2001wn},\cite{Pomeransky:2006bd},\cite{Elvang:2007rd},\cite{Evslin:2007fv},\cite{Iguchi:2007is},\cite{Elvang:2007hs} (to cite only a few). However, many of the properties of these exotic solutions remain to be analyzed, the reason for such being due to the fact that most of these space-times depend strongly on a large number of parameters. The study of the geodesic structure, possible extensions across the Killing horizons, causal relations and of properties such as regularity, stable causality or even global hyperbolicity of the \emph{domain of outer communications (d.o.c.)} seems in some cases to be unthinkable, despite the fact that such properties must be known for the space-time to have (at least) physical relevance. 

Recently a large amount of work has been done in this direction by \emph{Chru\'{s}ciel et al.} for singly and doubly spinning black rings \cite{Chrusciel:2008hg},\cite{Chrusciel:2009vr},\cite{Chrusciel:2010jg} and black saturn configurations \cite{Chrusciel:2010ix}. In particular, the maximal analytic extension of singly spinning black rings has been constructed in \cite{Chrusciel:2008hg} and moreover shown that the maximally extended space-time has all the nice properties mentioned above. Furthermore, it was demonstrated that the global structure of the singly spinning and the doubly spinning black ring resembles very closely that of the 4-dimensional Schwarzschild and Kerr black hole respectively \cite{Chrusciel:2008hg},\cite{Chrusciel:2009vr}.

In this paper we focus on a particular solution of Einstein-Maxwell's gravity, namely, that of a dipole black ring \cite{Emparan:2004wy}, i.e., a black ring charged under a purely magnetic gauge field, and study its global structure. It is expected, due to the close similarity to its neutral correspondent, that the overall properties characterizing this space-time will be somewhat reminiscent of those of the Reissner-Nordstr\"{o}m (RN) black hole . The key step will necessarily be that of constructing a global coordinate system capable of covering the entire extended space-time. Such extension should exist, at least in the non-extremal case, according to the general analysis done by \emph{Wald} and \emph{Rácz} \cite{Racz:1992bp}.

In general, given any black hole space-time, there exists a function $\Delta(y)$ for which at the horizon $y=y_{h},~\Delta(y_{h})=0$. Kruskal-Szekeres-type coordinates work well for non-extremal black hole space-times where $\Delta(y)$ exhibits a first order zero at $y=y_{h}$ and have been used to construct extensions in many situations \cite{Kruskal:1959vx},\cite{Graves:1960zz},\cite{Boyer:1967c},\cite{Carter:1968rr},\cite{Chrusciel:2008hg},\cite{Chrusciel:2009vr}.  However, in the extremal case, $\Delta(y)$ exhibits a second order zero at $y=y_{h}$ and the methods developed by \emph{Carter} for the extremal RN and Kerr metrics in \cite{Carter:1966zz},\cite{Carter:1966cz} seem to be more appropriate. It has also been constructed Kruskal-Szekeres-inspired coordinates for the extremal RN solution in \cite{Liberati:2000sq}. Unfortunately, both of these methods work well only in such situations because they can be reduced to a 2-dimensional Lorentzian problem. It should be noted that even in the case of the extremal Kerr metric, in order to use these methods one has to focus on the symmetry axis \cite{Carter:1966zz}. Thus, in more complicated cases, such as the one considered here, other methods have to be used.

This paper is organized as follows. In Sec.\ref{sdip} we give a general overview of the known properties of the dipole black ring space-time, which will be useful for the remaining parts of this work. In Sec.\ref{snex} we construct the maximal analytic extension of non-extremal dipole black rings using the methods developed in \cite{Chrusciel:2008hg},\cite{Chrusciel:2009vr} and study its causal structure. In Sec.\ref{sex} we focus on the extremal case and construct different extensions across the Killing horizons using the methods developed in \cite{Durkee:2008an},\cite{Kunduri:2007vf},\cite{Chrusciel:2009vr}. In Sec.\ref{hid}, motivated by the physical interpretation of one of these extensions we show the existence of a hidden symmetry associated with a conformal Killing tensor and a conformal Killing-Yano tensor in a lower dimensional space-time obtained by Kaluza-Klein reduction. Finally, in Sec.\ref{far} we give an overview of the results obtained here and pose some open problems.

\section{The dipole black ring space-time} \label{sdip}
The dipole black ring considered here is a solution of Einstein-Maxwell's equations derived from the action
\begin{equation} \label{act}
I=\frac{1}{16\pi G}\int d^5x\sqrt{-g}\left(R-\frac{1}{4}F_{\mu\nu}F^{\mu\nu}\right),
\end{equation}
where $F_{\mu\nu}$ is the usual Maxwell's stress-energy tensor.
It was first found in \cite{Emparan:2004wy} for a more general class of theories which includes a dilaton coupling and describes an asymptotically flat black hole with $S^1\times S^2$ horizon topology carying a dipole charge. In this section we briefly review some of the basic properties of the dipole black ring space-time which will be useful for what follows.

\subsection*{The metric}
The metric of the dipole black ring can be conveniently written in the form\footnote{If we denote the quantities in the metric given in \cite{Emparan:2004wy} with a bar and perform the coordinate transformation $\bar{x}=\frac{x-\lambda}{1-\lambda x}$, $\bar{y}=\frac{y-\lambda}{1-\lambda y}$, $(\bar{\phi},\bar{\psi})=\frac{(1-\lambda\nu)(1+\bar{\mu}\lambda)^{\frac{3}{2}})}{\sqrt{1-\lambda^2}}(\phi,\psi)$, $\bar{\nu}=\frac{\lambda-\nu}{1-\lambda\nu}$, $\bar{\lambda}=\lambda$, $\bar{\mu}=\frac{\mu-\lambda}{1-\lambda\mu}$, $\bar{R}=R\sqrt{\frac{1-\bar{\nu}\lambda}{(1+\bar{\mu}\lambda)^3}}$ we obtain the metric \eqref{ds1}}
\begin{equation} \label{ds1}
\begin{split}
ds^2&=-\frac{F(x)}{F(y)}\left(dt+RC(1+y)d\psi\right)^2 \\
&+\frac{R^2}{(x-y)^2}\left[-F(x)\left(\frac{G(y)}{F(y)}d\psi^2+\frac{F^2(y)}{G(y)}dy^2\right)+F^2(y)\left(\frac{F(x)}{G(x)}dx^2+\frac{G(x)}{F^2(x)}d\phi^2\right)\right],
\end{split}
\end{equation}
where $R$ is a constant, $C=\sqrt{\lambda\nu}$ and
\begin{equation} \label{fun}
F(\xi)=1-\mu\xi,~G(\xi)=(1-\xi^2)(1-\lambda\xi)(1-\nu\xi).
\end{equation}
The coordinates $t,y,x$ lie within the ranges $-\infty<t<\infty$, $-\infty< y <-1$, and $-1\le x\le 1$ respectively, while the periodicity of the angular coordinates $\psi,\phi$ is fixed by requiring regularity at the rotation axis. Moreover, the dimensionless parameters $\nu,\lambda,\mu$ must lie within the range $0<\nu\le\lambda\le\mu<1$.

The metric \eqref{ds1} together with the purely magnetic gauge field
\begin{equation} \label{gau}
A_{\phi}=\sqrt{3}C'R\frac{1+x}{F(x)}+k_{1},
\end{equation}
where
\begin{equation}
C'=\sqrt{(\mu-\lambda)(\mu-\nu)\frac{1-\mu}{1+\mu}},
\end{equation}
and $k_{1}$ is a constant, satisfy the equations of motion that arise by varying the action \eqref{act}.

\subsection*{Regularity at the axis $x=\pm1$}
At the axis $x=\pm1$, however, $G(x)=0$. So, in order to avoid conical singularities there the following condition must be imposed\footnote{The periodicities of the $\psi,\phi$ coordinates are thus $\Delta \psi=\Delta \phi=\frac{4\pi\sqrt{F(1)}}{|G'(1)|}$ with the condition \eqref{regcon} imposed.}:
\begin{equation} \label{regcon}
\left(\frac{1+\mu}{1-\mu}\right)^3=\left(\frac{1+\lambda}{1-\lambda}\right)^2\left(\frac{1+\nu}{1-\nu}\right)^2.
\end{equation}
In the case of the neutral singly spinning black ring \cite{Emparan:2001wn} one could avoid requiring a similar condition and instead obtain the 5-dimensional Myers-Perry family of solutions. Here it is manifest from the form of \eqref{ds1} that the choice $\mu=1$ would avoid $g_{\phi\phi}=0$ at $x=1$ but on the other hand $g_{\phi\phi}$ would become unbounded there and the solution would not be regular.

\subsection*{Asymptotic flatness}
Five-dimensional Minkowski space is reached when the 'point' $(x,y)\to(-1,-1)$ is approached. In the coordinates $(t,y,\psi,x,\phi)$ introduced above the metric is not manifestly asymptotically flat at this 'point', but in a similar fashion as in \cite{Elvang:2007hs},\cite{Durkee:2008an}, by performing the coordinate transformation
\begin{equation}
x=-1+\frac{2R^2}{\rho^2}(1+\lambda)(1+\nu)cos^2\theta,~y=-1-\frac{2R^2}{\rho^2}(1+\lambda)(1+\nu)sin^2\theta,
\end{equation}
one can bring the metric \eqref{ds1} at large $\rho$ into the flat space form
\begin{equation}
ds^2\approx-dt^2+d\rho^2+\rho^2(d\theta^2+cos^2\theta d\phi^2+sin^2\theta d\psi^2).
\end{equation}

\subsection*{Ergoregion}
The ergosurface lies at $y=-\infty$ where $F(y)\to+\infty$ and the metric in the coordinates presented in \eqref{ds1} is not analytic. However, it is straightforward to check that the transformation $y\to-\frac{1}{Y}$ extends the space-time smoothly across $Y=0$ to negative values of $Y$. In this new region one can introduce again $y$ by the inverse transformation and thus the surfaces $\{y=\pm\infty\}$ are identified. Therefore, from hereon we will take $y$ to lie within the interval $\{\frac{1}{\nu}<y<+\infty\}\cup\{-\infty<y\le-1\}$, keeping in mind that at $y=\pm\infty$ the coordinate $Y$ must be introduced. Thus, the ergoregion has $S^{1}\times S^{2}$ topology and is confined to the interval $\{\frac{1}{\nu}<y<+\infty\}$, where $\partial_t$ is spacelike and $F(y)<0$.

\subsection*{Killing horizons}
Even though the metric \eqref{ds1} resembles closely that of the singly spinning black ring, the function $G(y)$ has now two distinct first order zeros representing two different Killing horizons. The outer horizon is located at $y=\frac{1}{\nu}$, while the inner horizon sits at $y=\frac{1}{\lambda}$, which we denote by $y^\pm$. In the extremal case, when $\lambda=\nu$, the two horizons coincide and the function $G(y)$ acquires a second order zero.

Extensions across the Killing horizons, including the extremal case, can be easily constructed in a similar fashion as in \cite{Emparan:2001wn}, that is, by requiring the $g_{yy}$ component of the metric to vanish. In this case, an extension across the future event horizon can be attained by performing the transformation
\begin{equation} \label{cin}
dv=dt-RC(1+y)\frac{F(y)\sqrt{-F(y)}}{G(y)}dy,~d\hat{\psi}=d\psi+\frac{F(y)\sqrt{-F(y)}}{G(y)}dy.
\end{equation}
In this way, one can analytically continue across the surfaces $\{y=y^\pm\}$, with these surfaces being Killing horizons of the Killing vector fields,
\begin{equation} \label{av}
\xi^{\pm}=\partial_{v}+\Omega^{\pm}\partial_{\hat{\psi}},
\end{equation}
with angular velocities and surface gravities\footnote{The angular velocities given here are not canonically normalized. For a detailed analysis of the physical properties and asymptotic quantities of the dipole black ring see \cite{Emparan:2004wy}.}
\begin{equation} \label{sg}
\Omega^{\pm}=-\frac{1}{RC(1+y^{\pm})},~\kappa^{\pm}=\frac{1}{2y^{+}y^-RC}\frac{|y^{\pm}-y^{\mp}|(y^{\pm}-1)}{(\mu y^{\pm}-1)^{\frac{3}{2}}}.
\end{equation}

These extensions, however, are not maximal and resemble somewhat Eddington-Finkelstein-type coordinates but with the peculiarity that they are only valid in the ergoregion (as well as inside the horizon), and hence do not cover the full space-time parametrized by the coordinates of \eqref{ds1}. In Sec.\ref{sex}, we give a physical interpretation for this extension and moreover construct coordinates of the same type which do cover the original space-time entirely.

\subsection*{Neutral singly spinning limit}
Since the metric has been written in a slightly different form than usual it is worthwhile noting that the neutral singly spinning black ring can be obtained from \eqref{ds1} by setting $\mu=\lambda$ \footnote{Looking at the form of \eqref{gau}, the choice $\mu=\nu$ will also lead to the neutral case as long as one assumes a different hierarchy in the parameters, namely, $\lambda\le\nu$.}. In this case the gauge field \eqref{gau} becomes trivial and can be set to zero using the existent gauge freedom. To see this limit in the metric explicitly, we define the function
\begin{equation} \label{ghat}
\hat{G}(\xi)=\frac{G(\xi)}{F(\xi)},
\end{equation}
such that \eqref{ds1} takes exactly the same form as the original singly spinning black ring metric of \cite{Emparan:2001wn}. We will, from hereon, refer to this limit as the neutral limit. If now $\mu=\lambda$ we have that $\hat{G}(\xi)=(1-\xi^2)(1-\nu\xi)$ and the curvature singularity sits at $y=\frac{1}{\lambda}$, as in the neutral limit.

\subsection*{The curvature singularity}
The surface $\{y=\frac{1}{\mu}\}$ is timelike and represents a real curvature singularity. To show this explicitly note that if there exists a scalar invariant, constructed from the contraction of different copies of the Riemann tensor, that becomes unbounded as $y\to\frac{1}{\mu}$ then the surface $\{y=\frac{1}{\mu}\}$ is an unremovable curvature singularity. An example of such scalar is the Kretschmann scalar which reads
\begin{equation} \label{kret}
R_{\alpha\beta\gamma\delta}R^{\alpha\beta\gamma\delta}=\frac{12\left(G(\frac{1}{\mu})\right)^2(x-y)^4(1+\mathcal{O}(y-\frac{1}{\mu}))}{R^4\mu^6(\frac{1}{\mu}-x)^2(y-\frac{1}{\mu})^8},
\end{equation}
and is clearly unbounded as $y\to\frac{1}{\mu}$. Moreover, since that along every curve approaching the surface $\{y=\frac{1}{\mu}\}$ the Kretschmann scalar becomes unbounded, space-time is inextendible there.

\section{Maximal analytic extension of non-extremal dipole black rings}\label{snex}
In this section we construct a Kruskal-Szekeres-type extension for the non-extremal dipole black ring and study its causal structure.
The techniques employed here follow closely the work done in \cite{Chrusciel:2008hg},\cite{Chrusciel:2009vr} for different black ring configurations. The main results presented below will be based on the theorems and proofs resultant from a lengthly geodesic analysis carried out at the end of this section. 
\subsection{The extension}
To construct the maximal extension of this space-time we ought to construct Kruskal-Szekeres-type coordinates that are valid around each bifurcate Killing horizon at $y=y^{\pm}$. We note that in the non-extremal case $\nu\ne\lambda$ and hence $G(y)$, given in \eqref{fun}, has two first order zeros corresponding to each distinct horizon. The extension across each Killing horizon takes essentially the same form, so in what follows we denote any of the two horizons $y^{\pm}$ by $y_{h}$. 

We start by introducing new ingoing and outgoing coordinates $(v,u)$ in the same fashion as in \cite{Boyer:1967c}, i.e.
\begin{equation} \label{ct11}
\begin{split}
dv=&dt+\frac{\sigma}{y-y_{h}}dy,\\
du=&dt-\frac{\sigma}{y-y_{h}}dy,
\end{split}
\end{equation}
and further introduce a new angular coordinate $\hat{\psi}$ as
\begin{equation} \label{ct13}
d\hat{\psi}=d\psi-adt,
\end{equation}
where $\sigma$ and $a$ are constants to be adjusted by demanding regularity and analyticity of the new metric. In terms of the new coordinates $(v,u,\hat{\psi},\phi,x)$ the original coordinate differentials can be expressed as
\begin{equation} \label{diff1}
dt=\frac{du+dv}{2},~dy=\frac{H(y)}{2}(du-dv),~d\psi=d\hat{\psi}+a\frac{du+dv}{2},
\end{equation}
where we have defined
\begin{equation}
H(y)=\frac{y-y_{h}}{\sigma}.
\end{equation}
Furthermore, the metric coefficients in these new coordinates read
\begin{equation} \label{m1}
\begin{split}
g_{vv}=g_{uu}=&-\frac{F(x)}{4F(y)}\left(1+aRC(1+y)\right)^2-\frac{R^2F(x)F(y)}{2(x-y)^2}\left(\frac{a^2G(y)}{F^2(y)}+\frac{F(y)H^2(y)}{G(y)}\right),\\
g_{vu}=&-\frac{F(x)}{4F(y)}\left(1+aRC(1+y)\right)^2-\frac{R^2F(x)F(y)}{2(x-y)^2}\left(\frac{a^2G(y)}{F^2(y)}-\frac{F(y)H^2(y)}{G(y)}\right),\\
g_{v\hat{\psi}}=g_{u\hat{\psi}}=&-\frac{F(x)}{4F(y)}C(1+y)(1+aRC(1+y))-\frac{aR^2F(x)G(y)}{2(x-y)^2F(y)},\\ 
g_{\hat{\psi}\hat{\psi}}=&-\frac{F(x)}{F(y)}R^2C^2(1+y)^2-\frac{R^2F(x)G(y)}{(x-y)^2F(y)}.
\end{split}
\end{equation}
The Jacobian of the coordinate transformation \eqref{ct11} above is simply
\begin{equation}
\frac{\partial(v,u,\hat{\psi},\phi,x)}{\partial(t,y,\psi,\phi,x)}=-2H(y)^{-1},
\end{equation}
and hence the determinant of the metric in this new coordinate system reads
\begin{equation}\label{det1}
det(g_{v,u,\hat{\psi},\phi,x})=-\frac{F^2(x)F^4(y)H^2(y)}{4(x-y)^8}.
\end{equation}
Now, due to $H(y)$ the determinant has a second order zero at $y=y_{h}$ leading to a singular metric as $y\to y_{h}$. In order to remove this degeneracy we introduce new coordinates $(\hat{v},\hat{u})$ as in \cite{Kruskal:1959vx}
\begin{equation} \label{ct12}
\hat{v}=e^{\gamma v},~\hat{u}=e^{-\gamma u}.
\end{equation}
Thus, the coordinate differentials become
\begin{equation}
d\hat{v}=\gamma \hat{v}dv,~d\hat{u}=-\gamma \hat{u}du,
\end{equation}
while the Jacobian of this transformation reads
\begin{equation}
\frac{\partial(\hat{v},\hat{u},\hat{\psi},\phi,x)}{\partial(v,u,\hat{\psi},\phi,x)}=-\gamma^2\hat{v}\hat{u},
\end{equation}
leading to the determinant of the new metric in coordinates $(\hat{v},\hat{u},\hat{\psi},x,\phi)$
\begin{equation} \label{newdet}
det(g_{\hat{v},\hat{u},\hat{\psi},\phi,x})=-\frac{F^2(x)F^4(y)H^2(y)}{4\gamma^4(x-y)^8\hat{v}^2\hat{u}^2}.
\end{equation}
We now note that for $y>y_{h}$, the product $\hat{v}\hat{u}$ satisfies the relation
\begin{equation}
\hat{v}\hat{u}=e^{\gamma(v-u)}=\text{exp}\left(2\gamma\int^{y}\frac{\sigma}{y-y_{h}}dy\right)=e^{2\gamma ln(y-y_{h})}=(y-y_{h})^{2\gamma\sigma},
\end{equation}
therefore, in order to remove the degeneracy in the determinant \eqref{newdet} at $y=y_{h}$ we choose $2\gamma\sigma=1$ giving rise to $\hat{v}\hat{u}=(y-y_{h})=\sigma H(y)$. It now suffices to show that all the metric coefficients are analytic in the regions $\{\frac{1}{\mu}<y<y^+\}$ and  $\{y^-<y<+\infty\}\cup\{-\infty<y\le-1\}$ depending on $y_{h}$. The components of the metric tensor in the coordinate system $(\hat{v},\hat{u},\hat{\psi},\phi,x)$ are given by
\begin{equation} \label{m2}
g_{\hat{v}\hat{v}}=\frac{g_{vv}\hat{u}^2}{\gamma^2 \hat{u}^2\hat{v}^2},~g_{\hat{u}\hat{u}}=\frac{g_{uu}\hat{v}^2}{\gamma^2 \hat{u}^2\hat{v}^2},~g_{\hat{v}\hat{u}}=-\frac{g_{vu}}{\gamma^2 \hat{u}\hat{v}},~g_{\hat{v}\hat{\psi}}=\frac{g_{v\hat{\psi}}}{\gamma \hat{u}\hat{v} },~ g_{\hat{u}\hat{\psi}}=-\frac{g_{u\hat{\psi}}}{\gamma \hat{u}\hat{v} }.
\end{equation}
It is clear by looking at \eqref{m1} that such will be case if there is a multiplicative factor of $(y-y_{h})^2$ in the components $g_{vv},g_{uu}$ and a factor of $(y-y_{h})$ in $g_{v\hat{\psi}},g_{u\hat{\psi}}$ and in $g_{vu}$. This demands that  the following linear system of equations must be solved:
\begin{gather} \label{sys1}
1+aRC(1+y)=aRC(y-y_{h}),\\
\frac{a^2(1-y_{h}^2)(y^{\pm}-y_{h})}{y^{\pm}(1-\mu y_{h})^2}+\frac{y^{\pm}(1-\mu y_{h})}{\sigma^2(1-y_{h}^2)(y^{\pm}-y_{h})}=0.
\end{gather}
After some algebra we find that $a$ and $\sigma$ must take the values
\begin{equation} \label{crot}
a=-\frac{1}{RC(1+y_{h})},~\sigma=y^{+}y^{-}RC\frac{(\mu y_{h}-1)^{\frac{3}{2}}}{|y^{\pm}-y_{h}|(y_h-1)}.
\end{equation}
Comparing these expressions with \eqref{sg} we can identify $a=\Omega^{\pm}$ and $\gamma=\kappa^{\pm}$. This is a generic feature in the construction of Kruskal-Szekeres-type coordinates for non-extremal black hole space-times \cite{Kruskal:1959vx},\cite{Graves:1960zz},\cite{Boyer:1967c},\cite{Carter:1968rr},\cite{Chrusciel:2008hg},\cite{Chrusciel:2009vr}. We will now explain below how to build up the full extended space-time using these coordinates.

\newpage

\subsection{Causal structure}
\begin{wrapfigure}{r}{0.3\textwidth}
\vspace{-40pt}
  \begin{center}
    \includegraphics[width=0.35\textwidth, height=0.43\textheight]{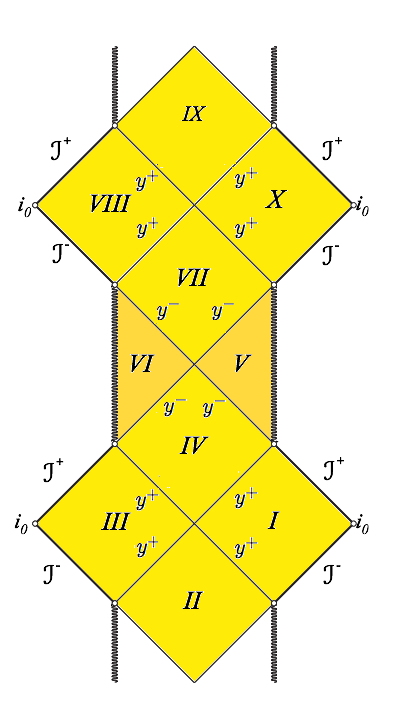}\label{fig:nex}
  \end{center}
  \vspace{-15pt}
  \caption{\small The causal structure of the non-extremal dipole black ring. This is not a conformal diagram nor the space-time is a product of the depicted diagram with $S^{1}\times S^{2}$ but it represents the causal relations between each region correctly.}
  \vspace{10pt}
    \label{fig1}
\end{wrapfigure}
The causal structure for the non-extremal dipole black ring is depicted in Fig.\ref{fig1} and resembles very closely that of the 4-dimensional Reissner-Nordstr\"{o}m black hole. Even though this diagram is the main result of the following sections it is instructive to present it beforehand for clarity of explanation. This picture has been constructed taking into consideration the causal properties described in the sections below and using the following iterative method. Suppose that we start in the region $\{y^+<y<+\infty\}\cup\{-\infty<y\le-1\}$ that we have denoted by $\mathcal{M}_{I}$ and introduce coordinates $(\hat{v}^{+},\hat{u}^{+},\hat{\psi}^{+})$ defined as above, where we have written the index $+$ to emphasize that we should take $y_{h}=y^{+}$. Then the above construction implies that we have produced an analytic Lorentzian metric on the set
\begin{equation}
\hat{\Omega}^{+}:=\left\{\hat{v}^{+},\hat{u}^{+}| - \frac{y^{+}+1}{2}\le\hat{v}^{+}\hat{u}^{+}<\frac{y^{+}-y^{-}}{y^{-}-1}\right\}\times S_{\hat{\psi}}^{1}\times S_{(x,\phi)}^{2},
\end{equation}
where the horizon $\{y=y^{+}\}$ sits at $\hat{v}^{+}\hat{u}^{+}=0$. It follows that the map
\begin{equation}\label{map}
(\hat{v}^{+},\hat{u}^{+},\hat{\psi}^{+},x,\phi)\to(-\hat{v}^{+},-\hat{u}^{+},\hat{\psi}^{+},x,-\phi),
\end{equation}
is an orientation-preserving analytic isometry of the analytically extended metric on $\hat{\Omega}^{+}$ and hence we can construct a manifold $\mathcal{M}_{1}$ obtained by gluing together two isometric copies of $(\mathcal{M}_{I},g^{+})$ and $\hat{\Omega}^{+}$ equipped with the same Lorentzian metric $g^{+}$ constructed above. We have denoted the second copy of $(\mathcal{M}_{I},g^{+})$ by $(\mathcal{M}_{III},g^{+})$, hence $\mathcal{M}_{1}=\mathcal{M}_{I}\cup\mathcal{M}_{II}\cup\mathcal{M}_{III}\cup\mathcal{M}_{IV}$ (see Fig.\ref{fig1}). The metric $g^{+}$ breaks down at $\hat{v}^{+}\hat{u}^{+}=\frac{y^{+}-y^{-}}{y^{-}-1}$, thus in the region $\mathcal{M}_{IV}$ where $\{\hat{v}^{+}>0,\hat{u}^{+}>0\}$ we introduce new coordinates $(\hat{v}^{-},\hat{u}^{-},\hat{\psi}^{-})$. This implies that the construction above has now produced another analytic Lorentzian metric on the set
\begin{equation}
\hat{\Omega}^{-}:=\left\{\hat{v}^{-},\hat{u}^{-}| - \frac{y^{-}-y^{+}}{1-y^{+}}<\hat{v}^{-}\hat{u}^{-}<\frac{y^{-}-\frac{1}{\mu}}{\frac{1}{\mu}-1}\right\}\times S_{\hat{\psi}}^{1}\times S_{(x,\phi)}^{2},
\end{equation}
where the horizon $y=y^{-}$ is now at $\hat{v}^{-}\hat{u}^{-}=0$. With the same map \eqref{map} but with the index $+$ exchanged with $-$ we can construct a manifold $\mathcal{M}_{2}$ equipped with the metric $g^{-}$ obtained by gluing together $\Omega^{-}$ and two isometric copies of $\mathcal{M}_{V}$, and hence $\mathcal{M}_{2}=\mathcal{M}_{IV}\cup\mathcal{M}_{V}\cup\mathcal{M}_{VI}\cup\mathcal{M}_{VII}$. Thus $\mathcal{M}_{2}$ overlaps with $\mathcal{M}_{1}$ in $\mathcal{M}_{IV}$ excluding its boundary. Similarly, in the region $\{\hat{v}^{-}>0,\hat{u}^{-}>0\}$ the metric $g^{-}$ breaks down at $y=y^{+}$ and we could introduce again $(\hat{v}^{+},\hat{u}^{+},\hat{\psi}^{+})$ coordinates obtaining a new patch $\mathcal{M}_{3}$. We label each overlapping patch as $\mathcal{M}_{n}$, where for $n$ odd we should introduce $(\hat{v}^{+},\hat{u}^{+},\hat{\psi}^{+})$ and for $n$ even $(\hat{v}^{-},\hat{u}^{-},\hat{\psi}^{-})$ coordinates. We name the space-time constructed by gluing together an infinite number of overlapping patches $\mathcal{M}_{n}$ as $(\mathcal{\tilde{M}},\tilde{g})$.

\subsection{Black hole and white hole regions}
With an extension across the Killing horizons $y=y^{\pm}$ one can easily show the existence of black hole and white hole regions when crossing the surfaces $\{y=y^{\pm}\}$. In order to do so, note that
\begin{equation} \label{hyp1}
g(\nabla_{y},\nabla_{y})=g^{yy}=-\frac{(x-y)^2G(y)}{R^2F(x)F(y)^2},
\end{equation}
is valid in either of the regions $\{y^+<y<+\infty\}\cup\{-\infty<y\le-1\}$ and $\{\frac{1}{\mu}<y<y-\}$ where $G(y)$ is negative and by analyticity it remains valid in the region $\{y^-<y<y^+\}$ where $G(y)$ is positive. Thus \eqref{hyp1} shows that the surfaces $\{y=y^\pm\}$ are null hypersurfaces with $y$ being a time function in the region $\{y^-<y<y^+\}$. If we now take any patch $\mathcal{M}_{n}$ with odd $n$, the usual choice of time orientation implies that along any future directed causal curve in the set $\{\hat{v}^{+}>0,\hat{u}^{+}>0\cup\hat{v}^{-}<0,\hat{u}^{-}<0\}$ $y$ is strictly decreasing and similarly in the set $\{\hat{v}^{+}<0,\hat{u}^{+}<0~\cup~\hat{v}^{-}>0,\hat{u}^{-}>0\}$ $y$ is strictly increasing. This means that, with respect to an observer in the asymptotic flat region, say $\mathcal{M}_{I}$, $B:=\{\hat{v}^{+}>0,\hat{u}^{+}>0\}$ is a black hole region since no future directed causal curve that crossed to $\mathcal{M}_{IV}$ can cross back to $\mathcal{M}_{I}$ as $y$ cannot increase along such curves, conversely, future directed causal curves are forced to leave the set $W:=\{\hat{v}^{+}<0,\hat{u}^{+}<0\}$ and hence these represent white hole regions. The arguments are similar for the regions covered by the patches $\mathcal{M}_{n}$ with even $n$.

\subsection{The topology of the event horizon}
As the metric \eqref{ds1} is regular everywhere in the \emph{d.o.c.} ($\mathcal{M}_{I}$) it is expected that the event horizon $\mathcal{H^+}$ exactly coincides with the Killing horizon at $y=y^+$. In order to show that this is exactly the case define the Killing vectors $K_{i},i=1,2,3$ as $\partial_{t},\partial_{\psi},\partial_{\phi}$ respectively. Then, one can compute the "area function" defined as the determinant of the matrix $g(K_{i},K_{j})$, which reads
\begin{equation} \label{af}
\frac{R^4G(x)G(y)}{(x-y)^4}.
\end{equation}
Now, since $x$ is confined to the region $-1\le x\le1$ then $G(x)$ is always positive, hence the sign of \eqref{af} depends only on $y$. Starting in the $d.o.c.$, where $G(y)$ is negative, one can easily see that at any point $p$ in this region the set of vectors spanned by the Killing vectors $K_{i}$ must be timelike. But, any event horizon crossing this region must be a null hypersurface invariant under all isometries where every Killing vector must be tangent to it. However, as we have stressed above, at each point we may construct a timelike Killing vector which by definition is not tangent to a null hypersurface. Therefore, as expected, the event horizon $\mathcal{H^+}$ must coincide with the Killing horizon at $y=y^+$ with topology $S^1\times S^2$. If we had instead started in $\mathcal{M}_{V}$ we could have proceeded similarly and shown that the event horizon $\mathcal{H^-}$ coincided with the Killing horizon at $y=y^-$.

\subsection{Maximality}
In this section we wish to show that the extension $(\tilde{\mathcal{M}},\tilde{g})$ is a maximal analytical extension of the space-time $(\mathcal{M},g)$. We start by recalling a useful definition: 

\begin{definition}A space-time $(\tilde{\mathcal{M}},\tilde{g})$ is said to be an extension of a space-time $(\mathcal{M},g)$ if there exists a map $\Psi: \mathcal{M}\to\tilde{\mathcal{M}}$ such that $\Psi^{*}\tilde{g}=g$ and $\Psi(\mathcal{M})\ne\tilde{\mathcal{M}}$, and is said to be maximal if no such extensions exist. 
\end{definition}

A natural way to prove maximality is then to demonstrate the inextendibility of a given space-time $(\tilde{\mathcal{M}},\tilde{g})$. The usual method of doing so is to show that every \emph{causal }geodesic $\gamma(s)$ within $(\tilde{\mathcal{M}},\tilde{g})$ is complete, meaning that it can be extended to infinite affine parameter $s$, or that it hits a real curvature singularity in finite affine time, beyond which space-time is inextendible. In fact, one can restrict the study of causal geodesics to timelike geodesics by first noting that the existence of a curvature singularity implies the existence of a scalar invariant that diverges as the singularity is approached by following any geodesic $\gamma(s)$ and second, that a scalar invariant, which remains unchanged under coordinate transformations, can be constructed using local geometric objects (an example of such is the Kretschmann scalar computed in \eqref{kret}). Then, one can show that the following proposition holds true (see \cite{Chrusciel:2008hg}, section 4.3):
\begin{proposition} \label{propm}
Suppose that every timelike geodesic $\gamma(s)$ in $(\tilde{\mathcal{M}},\tilde{g})$ is either complete, or some scalar invariant is unbounded on $\gamma(s)$. Then $(\tilde{\mathcal{M}},\tilde{g})$ is inextendible.
\end{proposition}

With this in mind, we show in Sec.\ref{geo} below that every timelike geodesic $\gamma(s)$ is complete within $(\tilde{\mathcal{M}},\tilde{g})$. This is due to the fact that, since in the region $\{\frac{1}{\mu}<y<y^-\}$ $\partial_t$ is a timelike Killing vector field, the curvature singularity $\{y=\frac{1}{\mu}\}$ is a timelike hypersurface acting as a repellent to timelike geodesics, but not to null and spacelike geodesics nor non-geodesic curves. Thus $(\tilde{\mathcal{M}},\tilde{g})$ is a timelike complete but not a causal geodesically complete space-time. This particular feature had already been observed in the 4-dimensional Reissner-Nordstr\"{o}m black hole \cite{Graves:1960zz},\cite{Hawking:ls}. This then allows us to write down the following theorem:
\begin{theorem} \label{tmaxi}
$(\tilde{\mathcal{M}},\tilde{g})$ is a maximal analytical extension of the non-extremal dipole black ring space-time $(\mathcal{M}_{I},g)$.
\end{theorem}

\subsection{Non-global hyperbolicity}
It is a general feature of space-times with a similar causal structure as the one presented above to be non-globally hyperbolic. This is because any two consecutive copies of $\mathcal{M}_{n}$ for odd $n$ are separated by a copy of  $\mathcal{M}_{n}$ with even $n$ which contains a white whole region. Then, since each copy of $\mathcal{M}_{n}$ for odd $n$ contains an asymptotically flat region, and hence geodesics which are fully contained in it, every space-like surface one could imagine defining on $\mathcal{M}_{n_1}$ with $n_1$ odd will not be crossed by the geodesics contained in the asymptotically flat region of any other $\mathcal{M}_{n}$. However, we can show global hyperbolicity if we restrict ourselves to the \emph{d.o.c.}, to be more precise, one can show that:
\begin{proposition}
The space-time given by any copy of $(\mathcal{M}_{n},g^{+})$ for odd $n$ is globally hyperbolic.
\end{proposition}
The proof of this statement readily follows from the global hyperbolicity of the neutral limit which was proven in (\cite{Chrusciel:2008hg}, section 4.4). The difference being that now $G(\xi)$ in \cite{Chrusciel:2008hg} should be replaced by $\hat{G}(\xi)$, as given in \eqref{ghat}, and that in the regions $\mathcal{M}_{II}, \mathcal{M}_{IV}$ null geodesics instead of hitting the curvature singularity are now forced to leave these regions to the past and to the future respectively according to the geodesic analysis carried out below, never to return, since $\dot{y}$ is a time function there.

\subsection{Geodesic analysis} \label{geo}
In this section we wish to show that causal geodesics parametrized by
\begin{equation} \label{cur}
\gamma(s)=(t(s),y(s),x(s),\psi(s),\phi(s)),
\end{equation}
can be infinitely extended in their affine parameter $s$ within the space-times $(\tilde{\mathcal{M}},\tilde{g})$ constructed above or hit the curvature singularity, more precisely, we ought to prove the following theorem:
\begin{theorem} \label{tmax}
All maximally extended causal  geodesics in $(\tilde{\mathcal{M}},\tilde{g})$ are either complete or reach the singular boundary $\{y=\frac{1}{\mu}\}$ in finite affine time.
\end{theorem}
We will show that this theorem holds true for non-extremal dipole black rings by considering the geodesic equations obtained from the Lagrangian
\begin{equation} \label{lag2}
\mathcal{L}(x^{\mu},\dot{\gamma}^{\nu})=\frac{1}{2}g(\dot{\gamma},\dot{\gamma}),
\end{equation}
where the 'dot' represents a derivative with respect to $s$, and the constants of motion
\begin{equation}
\mathcal{L}:=g(\dot{\gamma},\dot{\gamma}),~ p_{t}:=g(\partial_{t},\dot{\gamma}),~p_{\psi}:=g(\partial_{\psi},\dot{\gamma}),~p_{\phi}:=g(\partial_{\phi},\dot{\gamma}),
\end{equation}
from which we can write down the following identities:
\begin{equation} \label{cpsi}
\dot{\psi}=\frac{(x-y)^2F(y)}{R^2F(x)G(y)}(RC(1+y)p_{t}-p_{\psi}),
\end{equation}
\begin{equation} \label{ttt}
\dot{t}=-\frac{F(y)}{F(x)}p_{t}-RC(1+y)\frac{(x-y)^2F(y)}{R^2F(x)G(y)}(RC(1+y)p_{t}-p_{\psi}),
\end{equation}
\begin{equation} \label{cphi}
\dot{\phi}=\frac{(x-y)^2F^2(x)}{R^2F^2(y)G(x)}p_{\phi},
\end{equation}
\begin{equation} \label{lag}
\begin{split}
\mathcal{L}=-\frac{F(y)}{F(x)}p_{t}^2-\frac{R^2F^2(y)F(x)}{(x-y)^2G(y)}\dot{y}^2&-\frac{(x-y)^2F(y)}{R^2F(x)G(y)}(RC(1+y)p_{t}-p_{\psi})^2\\
&+\frac{R^2F^2(y)F(x)}{(x-y)^2G(x)}\dot{x}^2+\frac{(x-y)^2F^2(x)}{R^2G(x)F^2(y)}p_{\phi}^2,
\end{split}
\end{equation}
where we have rescaled $\mathcal{L}\to2\mathcal{L}$ to avoid useless factors of $2$.

Now, due to the similarity of the form of the metric \eqref{ds1} with the neutral case, the analysis of the geodesics of the non-extremal dipole black ring is very similar to that of the neutral black ring, which have been extensively studied in \cite{Chrusciel:2008hg}. In fact, if we split up the $y$ axis into different intervals, it will only be in the region $\{\frac{1}{\mu}\le y<y^{-}\}$ that the arguments of \cite{Chrusciel:2008hg} cannot be modified to include this case. Therefore, we will only outline the main arguments in the remaining intervals and ask the reader to check \cite{Chrusciel:2008hg} for a more detailed analysis, while the interval $\{\frac{1}{\mu}\le y<y^{-}\}$ will be analyzed carefully here. We will be mainly concerned with timelike geodesics but some of the results presented here will also concern any causal geodesic. Moreover, our analysis will only be restricted to the two patches $\mathcal{M}_{1},\mathcal{M}_{2}$, since as explained above, the entire space-time can be covered by infinitely many copies of these regions. 

\subsubsection*{Geodesics in the region $\{(y^+ + \epsilon)<y<+\infty\}\cup\{-\infty<y\le-1\}$}
In this interval we have that $G(y)$ is negative while $F(y)$ is positive for negative $y$ and negative for positive $y$. We further assume that $\epsilon>0$. We wish to show that the following statement is true:
\begin{proposition} \label{pdoc}
For any $\epsilon>0$, maximally extended geodesics within the region $\{(y^{+}+\epsilon)\le y\le-1\}$ are either complete, or acquire a smooth endpoint at $\{y=y^{+}+\epsilon\}$.
\end{proposition}
If we first consider geodesics that approach the asymptotically flat region, that is, $\lim_{s\to\infty}(x(s)-y(s))=0$, then such geodesics remain in that region and are maximally extended there. Therefore assume that $|x-y|>0$ and consider the geodesic segment $\{-2\le\ y(s)\le -1\}$. In this case we can introduce well behaved polar type coordinate near the axes of rotation $G(y)=0$ and $G(x)=0$ as
\begin{equation} \label{deft}
d\theta=\frac{dx}{\sqrt{G(x)}},~d\omega=\frac{dy}{\sqrt{G(y)}},
\end{equation}
and then rewrite \eqref{lag} as
\begin{equation}\label{pppp}
\begin{split}
\frac{(x-y)^2}{R^2F(y)}\left(\mathcal{L}+\frac{F(y)}{F(x)}p_{t}^2\right)=&F(x)F(y)(\dot{\omega}^2+\dot{\theta}^2)+\frac{(x-y)^4F(x)}{R^4G(x)F^3(y)}p_{\phi}^2\\
&+\frac{(x-y)^4}{R^4F(x)|G(y)|}(RC(1+y)p_{t}-p_{\psi})^2.
\end{split}
\end{equation}
Now, since the coefficients of the two first terms on the RHS of Eq.\eqref{pppp} are bounded from above and away from zero and since the LHS is also bounded, we have a bound on $\dot{\omega}^2+\dot{\theta}^2$. Moreover a non-zero $p_{\phi}$ prevents $x$ from approaching $-1$, unless $x-y\to0$, similarly, a non-zero $p_{\psi}$ prevents $y$ from approaching $-1$. Also, from \eqref{cpsi}-\eqref{cphi} we can find bounds on $|\dot{\psi}|,|\dot{\phi}|$ when $p_{\phi}$ or $p_{\psi}$ are non-zero and from \eqref{pppp} otherwise. This implies that there exists a constant $C_{1}$ such that
\begin{equation}
\dot{t}^2+\dot{\theta}^2+\dot{\omega}^2+\dot{\psi}^2+\dot{\phi}^2\le C_{1}.
\end{equation}
Furthermore, taking now the segment $\{y^+<y<+\infty\}\cup\{-\infty<y\le-2\}$, making the transformation $Y=-\frac{1}{y}$ as explained in Sec.\ref{sdip}, such that we have an analytic metric through the ergosurface, and proceeding similarly we find that
\begin{equation}
\dot{t}^2+\dot{\theta}^2+\dot{Y}^2+\dot{\psi}^2+\dot{\phi}^2\le C_{2},
\end{equation}
and therefore those geodesics in this region obey the properties of Prop.\ref{pdoc}.

\subsubsection*{Geodesics in the region $\{y^{-}< y(s)< y^+\}$}
In this region we have that $G(y)$ is positive while $F(y)$ is negative, thus we rewrite \eqref{lag} as
\begin{equation} \label{lagin}
F^2(y)\left(\frac{\dot{y}^2}{G(y)}-\dot{\theta}^2\right)=\Theta +\frac{(x-y)^4F(x)}{R^2G(x)F^2(y)}p_{\phi}^2,
\end{equation}
where
\begin{equation}
\Theta=\frac{(x-y)^2}{R^2F(x)}\left(-\mathcal{L}+\frac{|F(y)|}{F(x)}p_{t}^2\right)+\frac{(x-y)^4|F(y)|}{R^2F^2(x)G(y)}(RC(1+y)p_{t}-p_{\psi})^2.
\end{equation}
Our goal here is to show the veracity of the proposition below:
\begin{proposition} \label{pinout}
Maximally extended causal geodesics in the region $\{y^{-}< y(s)< y^+\}$ reach the bifurcate Killing horizons $\{\hat{v}^\pm\hat{u}^\pm=0\}$ in finite affine time and are smoothly extendible there.
\end{proposition}
Now, as explained in the discussion around \eqref{hyp1}, $y$ is a time function in this region and strictly decreasing for future directed causal curves, thus geodesics entering this region are directed towards the inner horizon $y=y^{-}$. Also, since the RHS of \eqref{lagin} is positive, we have that
\begin{equation}
\frac{R^2F^2(y)F(x)}{(x-y)^2G(y)}\dot{y}^2\ge-\mathcal{L}.
\end{equation}
Choosing proper time as the parameter along $\gamma(s)$ for timelike geodesics we obtain $\mathcal{L}=-2$ (since we have sent $\mathcal{L}\to2\mathcal{L}$) and thus the proper length $L$ along the curve $\gamma(s)$ satisfies the following relation for some constant $\varepsilon$:
\begin{equation}
L=\int_{y^+}^{y^-}\left|\frac{ds}{dy}\right|dy\le\frac{1}{\varepsilon}\int_{y^+}^{y^-}\frac{|F(y)|}{\sqrt{G(y)}}dy<\infty.
\end{equation}
Hence, future directed timelike geodesics after crossing $y=y^{+}$ reach $y=y^{-}$ in finite proper time except if $\dot{y}$ and $\dot{\theta}$ become unbounded as $y\to y^\pm$. One then needs to show that in fact $\dot{y}$, $\dot{\theta}$ are bounded in this region and moreover, to get extendibility across the Killing horizons, that the remaining components $\dot{\gamma}^{\mu}$ are also bounded in the appropriate coordinate system. 

In order to adapt the arguments of \cite{Chrusciel:2008hg} to the case under consideration one needs to split the set $\{y^{-}< y(s)< y^+\}$ into two sets as $\{y^{-}+\epsilon\le y(s)< y^+\}$ and $\{y^{-}< y(s)\le y^{+}-\epsilon\}$ for some $\epsilon>0$ since none of the coordinate systems $(\hat{v}^{\pm},\hat{u}^{\pm})$ is simultaneously valid at both surfaces $\{y=y^\pm\}$. The analysis is the same for both so we focus on the set $\{y^{-}+\epsilon\le y(s)< y^+\}$. Here, we introduce the coordinates $(\hat{v}^{+},\hat{u}^{+},\hat{\psi}^{+})$ and from the relations \eqref{ct12}, \eqref{ct11} and \eqref{ct13} we have that as $y\to y^+$
\begin{equation} \label{noz}
\begin{split}
\frac{d\hat{v}^+}{ds}=&-\gamma\left(\hat{v}\frac{F(y)}{F(x)}p_t + \frac{y^- y^+ RCF(y)}{\hat{w}(1-y^+)(y-y^-)}(\beta(x,y^+)+\alpha(x,y^+))\right),\\
\frac{d\hat{u}^+}{ds}=&-\gamma\left(\hat{w}\frac{F(y)}{F(x)}p_t + \frac{y^- y^+ RCF(y)}{\hat{v}(1-y^+)(y-y^-)}(\beta(x,y^+)-\alpha(x,y^+))\right),\\
\frac{d\hat{\psi}^+}{ds}=&-\frac{(x-y)^2F(y)(y-y^+)}{F(x)G(y)(1+y^+)}(RC(1+y)p_t-p_\psi)+\frac{F(y)}{RC(1+y^+)F(x)}p_t ,
\end{split}
\end{equation}
where we have defined the functions $\alpha(x,y)$ and $\beta(x,y)$ as
\begin{equation} \label{albe}
\beta(x,y):=\frac{(x-y)^2(RC(1+y)p_{t}-p_{\psi})}{R^2F(x)},~\alpha(x,y):=\sqrt{-F(y)}\dot{y}
\end{equation}
Now, looking at \eqref{noz} we see that $\dot{\hat{\psi}}^+$ is well behaved at $y=y^+$ since the factor of $(y-y^+)$ in the numerator cancels out the one coming from $G(y)$. On the other hand, $\dot{\hat{v}}^+$ and $\dot{\hat{u}}^+$ may diverge as $y\to y^+$ since $\hat{v}^+\hat{u}^+=0$ there. So we proceed by noting that the functions $\alpha(x,y)$ and $\beta(x,y)$ appear naturally in the Lagrangean, to see this we rewrite \eqref{lag} explicitly as
\begin{equation} \label{vbou1}
-\frac{R^2F(x)F(y)}{(x-y)^2G(y)}(\alpha^2(x,y)-\beta^2(x,y))=-\mathcal{L}-\frac{F(y)}{F(x)}p_{t}^2 + \frac{R^2F^2(y)}{(x-y)^2}\dot{\theta}^2+\frac{(x-y)^2F^2(x)}{R^2G(x)F^2(y)}p_{\phi}^2.
\end{equation}
The aim now is to show that the LHS of the equation above has a limit as $y\to y^+$ and furthermore that from here $\dot{\hat{v}}$ and $\dot{\hat{u}}$ also have a limit as $y\to y^+$. To do this we first note that in the region under consideration $y$ is a time function as it is decreasing along any future directed causal geodesic and thus $\dot{y}$ has a constant sign. This means that $y$ can be used as a parameter along the geodesics. Hence, by considering the Euler-Lagrange equations for $y$ derived from the Lagrangean \eqref{lag2} and taking $y$ as the parameter along such curves, one can find after a series of manipulations (see \cite{Chrusciel:2008hg}) the relation
\begin{equation}\label{eqh}
\frac{2}{\sqrt{|G(y)|}}\frac{d}{dy}\left(\frac{R^2F(x)\sqrt{|F^3(y)|}}{(x-y)^2\sqrt{\hat{G}(y)}}\dot{y}\right)=\pm\Xi\frac{ds}{dy},
\end{equation}
where we have used the definition of $\hat{G}(\xi)$ given in \eqref{ghat}, and
\begin{equation} \label{eqlo1}
\begin{split}
\Xi=&\frac{F'(y)}{F(x)}p_t^2+2\frac{C(x-y)^2}{RF(x)\hat{G}(y)}(RC(1+y)p_t-p_\psi)p_t\\
&-\frac{\partial}{\partial y}\left(\frac{\hat{G}(y)}{(x-y)^2}\right)\frac{(x-y)^4}{R^2F(x)\hat{G}^2(y)}(RC(1+y)p_t-p_\psi)^2\\
&+\frac{\partial}{\partial y}\left(\frac{R^2F^2(y)}{(x-y)^2}\right)\left[\frac{(x-y)^4}{R^4F(x)F^2(y)\hat{G}(y)}(RC(1+y)p_t-p_\psi)^2  +\frac{(x-y)^2}{R^2F^2(y)}\left(\mathcal{L}+\frac{F(y)}{F(x)}p_t^2\right) \right].
\end{split}
\end{equation}
Note that Eq.\eqref{eqh} holds whatever the sign of $F(y)G(y)$ might be. In the region of current interest we take the $-$ sign on the RHS.
If we now define
\begin{equation} \label{eqh3}
\hat{h}:=\frac{R^2F(x)\sqrt{-F^3(y)}}{(x-y)^2\sqrt{\hat{G}(y)}}\dot{y},
\end{equation}
then \eqref{eqh} takes the form
\begin{equation} \label{eqh2}
\frac{d\hat{h}^2}{dy}=\pm\frac{R^2F(x)F^2(y)}{(x-y)^2}\Xi.
\end{equation}
We then have two cases to consider: \\ \\
\textbf{Case 1:} $RC(1+y^+)p_t-p_\psi=0$\\
If this condition is imposed then by looking at \eqref{eqlo1} we get that the RHS of \eqref{eqh2} is bounded. Using the relation between $\dot{y}$ and $\hat{h}$, i.e., Eq.\eqref{eqh3} we obtain a bound on $|\dot{y}|$ such that
\begin{equation} \label{zbou1}
|\dot{y}|\le C_{3}\sqrt{-G(y)}.
\end{equation}
Inserting this into Eq.\eqref{lagin} we obtain
\begin{equation}
\dot{\theta}^2+\frac{p_\phi^2}{G(x)}\le C_{4},
\end{equation}
hence, if $p_\phi\ne0$ then $x$ never approaches $\pm1$ and thus if $p_\phi=0$ or otherwise we find a bound on $|\dot{\theta}|$. Moreover plugging in \eqref{zbou1} into \eqref{noz} we find that
\begin{equation}
\frac{d(ln(\hat{u}^+/\hat{v}^+))}{dy}\le\frac{C_{5}}{\sqrt{|G(y)|}},
\end{equation}
where $C_{5}$ vanishes if $p_t=0$. From here we obtain $\hat{v}^+=\rho(y)\hat{u}^{+}$ for some function $\rho(y)$ which has a finite limit as $y\to y^+$. Feeding this back into \eqref{noz}, and since this is also valid in set $\{y^{-}< y(s)\le y^{+}-\epsilon\}$, we get
\begin{equation}
|\dot{\hat{v}}^{\pm}| +|\dot{\hat{u}}^\pm| \le C_{6},
\end{equation}
and hence smooth extendibility of the geodesics at $y=y^+$ follows. \\ \\
\textbf{Case 2:} $RC(1+y^+)p_t-p_\psi\ne0$\\
In this case looking at Eq.\eqref{eqlo1} we see that the most singular term as $y\to y^+$ is the third term on the RHS, hence using \eqref{eqh} we obtain,
\begin{equation}
\frac{d\hat{h}^2}{dy}=\frac{y^+|F^3(y^+)|(RC(1+y^+)p_t-p_\psi)^2}{(y-y^-)(y^+-1)(y^++1)(y-y^+)^2} + \mathcal{O}((y-y^+)^{-1}),
\end{equation}
which by integration leads to
\begin{equation}
\hat{h}^2=\frac{y^+|F^3(y^+)|(RC(1+y^+)p_t-p_\psi)^2}{(y-y^-)(y^+-1)(y^++1)(y-y^+)} + \mathcal{O}(ln|y-y^+|).
\end{equation}
Then, from Eq.\eqref{eqh3} we find that,
\begin{equation} \label{eqy1}
\dot{y}=\frac{(x^+-y^+)^2}{R^2\sqrt{|F(y^+)|}F(x^+)}(RC(1+y^+)p_t-p_\psi)+\mathcal{O}((y-y^+)ln|y-y^+|).
\end{equation}
where $x^+:=\lim_{y\to y^+}x(y)$, and hence $|\dot{y}|$ is bounded. 

We now need to know how $x^+$ is attained so that a bound on $\dot{\theta}$ can be found. From \eqref{lagin} multiplied by $(ds/dy)^2$ we get that
\begin{equation}
\left|\frac{d\theta}{dy}\right|\le \sqrt{\frac{F(x)}{|G(y)|}},
\end{equation}
which integrating gives $|\theta(z)-\theta(y^+)|\le C_{7}\sqrt{|y-y^+|}$. Hence, from the definition of $\theta$, Eq.\eqref{deft}, we find
\begin{equation}
|x(y)-x(y^+)|\le C_{8}\begin{cases}
|y-y^+|, &\text{if~$G(y^+)=0$,}\\
\sqrt{|y-y^+|},&\text{otherwise.}
\end{cases}
\end{equation}
Inserting this into \eqref{lagin} we obtain
\begin{equation} \label{tz2}
\dot{\theta}^2+\frac{p_\phi^2}{G(x)}\le \frac{C_{9}}{\sqrt{|y-y^+|}}.
\end{equation}
To show that $\dot{\theta}$ does not diverge at $y=y^+$ we further need the Euler-Lagrange equation for $x$. As in the $y$ case, after a series of rearrangements we find
\begin{equation} \label{tz1}
\frac{2}{\sqrt{G(x)}}\frac{d}{ds}\left(\frac{R^2F^2(y)F(x)}{(x-y)^2}\dot{\theta}\right)=\hat{\Xi},
\end{equation}
where
\begin{equation}
\begin{split}
\hat{\Xi}=&\frac{(x-y)^2}{R^4\hat{G}(x)F^2(y)}p_\phi^2\left[-\frac{\partial}{\partial x}\left(\frac{F(x)}{(x-y)^2}\right)+\frac{F^2(x)}{\hat{G}(x)}\frac{\partial}{\partial x}\left(\frac{\hat{G}(x)}{F(x)(x-y)^2}\right)\right]\\
&-\frac{F'(x)F(y)}{F^2(x)}p_t^2+\frac{\partial}{\partial x}\left(\frac{F(x)}{(x-y)^2}\right)\frac{(x-y)^2}{F(x)}\left(\mathcal{L}+\frac{F(y)}{F(x)}p_t^2\right).
\end{split}
\end{equation}
Now, multiplying \eqref{tz1} by $ds/dy$ and using Eq.\eqref{tz2} we find (see \cite{Chrusciel:2008hg})
\begin{equation}
\frac{d}{dy}\left(\frac{R^2F^2(y)F(x)}{(x-y)^2}\dot{\theta}\right)=\mathcal{O}(|y-y^+|^{-\frac{3}{4}}),
\end{equation}
hence, since the RHS is integrable then $|\dot{\theta}|$ is bounded as $y\to y^+$. It remains to show that $|\dot{\hat{v}}^+|$ and $|\dot{\hat{u}}^+|$ are bounded. Taking Eq.\eqref{vbou1} and multiplying it by $|G(y)|$ we find that 
\begin{equation} \label{lim1}
\lim_{y\to y^+}\alpha(x,y)=\pm\lim_{y\to y^+}\beta(x,y),
\end{equation}
and moreover $\lim_{y\to y^+}\beta(x,y)\ne0$ since $RC(1+y^+)p_t-p_\psi\ne0$. So suppose first that this holds with the $+$ sign. We can write $[\alpha(x,y)-\beta(x,y)]/G(y)$ as $[(\alpha^2(x,y)-\beta^2(x,y)])/G(y)/[\alpha(x,y)+\beta(x,y)]$ and again from Eq.\eqref{vbou1} we find that the limit
\begin{equation} \label{limb1}
\lim_{y\to y^+}\frac{\alpha(x,y)-\beta(x,y)}{G(y)}
\end{equation}
exists, thus we can write
\begin{equation}
\dot{\hat{u}}^+=\rho(s)\hat{u}^+,
\end{equation}
for some function $\rho(s)$. By integration $\hat{u}^+$ has a non-zero limit as $y\to y^+$ and thus $|\dot{\hat{u}}^+|$ is bounded. Moreover, since $\hat{v}^+\hat{u}^+=0$ at the horizon then we must have $\hat{v}^+=0$ and Eq.\eqref{vbou1} shows that $|\dot{\hat{v}}^+|$ is also bounded. Thus, since this analysis is valid as well for the set $\{y^{-}< y(s)\le y^{+}-\epsilon\}$, we have that in general
\begin{equation}
\hat{v}^\pm + \hat{u}^{\pm} + |\dot{\hat{v}}^+| + |\dot{\hat{u}}^+| + |\dot{\hat{\psi}}|+|\dot{y}| + |\dot{x}|  \le C_{10}.
\end{equation}
Similar arguments apply if we had taken the $-$ sign in \eqref{lim1}. This completes the proof of Prop.\ref{pinout}.

\subsubsection*{Geodesics in the region $\{\frac{1}{\mu}< y(s)\le y^{-}-\epsilon\}$}
In this region we have that both $G(y)$ and $F(y)$ are negative and we assume that $\epsilon>0$. Then equation \eqref{lag} can be written as
\begin{equation} \label{tc}
\begin{split}
\mathcal{L}-\frac{|F(y)|}{F(x)}p_{t}^2+\frac{(x-y)^2|F(y)|}{R^2F(x)|G(y)|}(RC(1+y)p_{t}-p_{\psi})^2=&\frac{R^2F^2(y)F(x)}{(x-y)^2|G(y)|}\dot{y}^2\\
&+\frac{R^2F^2(y)F(x)}{(x-y)^2G(x)}\dot{x}^2+\frac{(x-y)^2F^2(x)}{R^2G(x)F^2(y)}p_{\phi}^2.
\end{split}
\end{equation}
We now want to show the following proposition:
\begin{proposition} \label{ptc}
Maximally extended causal geodesics in the region $\{\frac{1}{\mu}\le y\le y^{-}-\epsilon\}$, for $\epsilon>0$, are either complete (or reach $\{y=\frac{1}{\mu}\}$ in finite affine time) or acquire a smooth endpoint at $\{y=y^{-}-\epsilon\}$.
\end{proposition}
First, assume that geodesics can in fact reach $y=\frac{1}{\mu}$, then the last two terms on the LHS of \eqref{tc} necessarily vanish as $y\to\frac{1}{\mu}$ since $F(y)$ vanishes at that point, leading to the behavior
\begin{equation}
\mathcal{L}\to\frac{R^2F(y)^2F(x)}{(x-y)^2|G(y)|}\dot{y}^2+\frac{R^2F(y)^2F(x)}{(x-y)^2G(x)}\dot{x}^2+\frac{(x-y)^2F(x)^2}{R^2G(x)F(y)^2}p_{\phi}^2,~\text{as}~y\to\frac{1}{\mu}.
\end{equation}
From here we can conclude two things, since the RHS is positive and the LHS is constant and as we approach the singularity the last term on RHS diverges we must have that $p_{\phi}=0$ for any geodesic reaching $y=\frac{1}{\mu}$. Moreover, since $\mathcal{L}<0$ for timelike geodesics and the RHS is positive then we conclude that timelike geodesics cannot reach the singularity. Hence we can write down the following lemma:
\begin{lemma} \label{lem1}
Causal geodesics reaching the singularity must be null and have $p_\phi=0$.
\end{lemma}
We now want to show that a subset of this special class of null geodesics reaches ${y=\frac{1}{\mu}}$ in finite affine time. For this we rewrite \eqref{tc} with $\mathcal{L}=0$ and $p_\phi=0$ as
\begin{equation} \label{tc1}
\frac{R^2F^2(y)F(x)}{(x-y)^2|G(y)|}\dot{y}^2=-\frac{|F(y)|}{F(x)}p_{t}^2-\frac{R^2F^2(y)F(x)}{(x-y)^2}\dot{\theta}^2+\frac{(x-y)^2|F(y)|}{R^2F(x)|G(y)|}(RC(1+y)p_{t}-p_{\psi})^2.
\end{equation}
Thus, since the first two terms on the RHS are negative and the LHS is positive, there exists a constant $C_{11}>0$ such that
\begin{equation} \label{fini}
\dot{y}\le-\frac{C_{11}}{\sqrt{|F(y)|}},
\end{equation}
where we have chosen the $-$ sign because we are interested in geodesics directed towards the singularity. Choosing $s$ as an affine parameter along these curves we arrive at
\begin{equation} \label{fini1}
\Delta s\le -\frac{1}{C_{11}}\int_{y^-}^{\frac{1}{\mu}}\sqrt{|F(y)|}dy<\infty.
\end{equation}
Hence, these geodesics hit the singularity in finite affine time except if $\dot{y}$ and $\dot{\theta}$ become unbounded before reaching that set. We will see below that this possibility cannot occur. However, when writing Eq.\eqref{fini} and Eq.\eqref{fini1} we have assumed two things. Firstly, that neither $p_t$ nor $p_\psi$ vanish simultaneously and secondly, that $\dot{y}$ has a constant negative sign, or equivalently, that there are no turning points where $\dot{y}=0$ for some $y_0\in(\frac{1}{\mu}+\varepsilon,y^{-}-\varepsilon)$ with small $\varepsilon>0$. With respect to the first assumption, due to the geodesic analysis which will be carried out in Sec.\ref{f0z} for zero energy ($p_t=0$) null geodesics, we have that if $p_\phi=p_\psi=0$ then from Eq.\ref{hj3} it necessarily implies $\dot{x}=\dot{y}=0$, and moreover from Eqs.\eqref{cpsi}-\eqref{cphi}, $\dot{\gamma}^\mu=0$, so we can ignore this case. With respect to the turning points, it is a hard task, in general, to deduce their existence, nevertheless, in the case $p_\phi=p_t=0$ and $p_\psi\ne0$ one can check, from the analysis of Sec.\ref{f0z} (see Eq.\eqref{hj4}), that the only turning point is at the ergosurface $\{y=\pm\infty\}$ if $c=0$, in fact, for these geodesics we have that $\dot{\theta}$ is always bounded and equal to zero. Therefore, in the region of current interest there are no turning points when $p_\phi=p_t=c=0,p_\psi\ne0$ and thus these geodesics reach the singularity in finite affine time\footnote{In fact one can precisely calculate the proper time ($\Delta s$) for such geodesics to hit the singularity. For this we take \eqref{hj4} and integrate it for a fixed $x=x_{0}$ from the ergosurface sitting at $y=+\infty$ to the curvature singularity at $y=\frac{1}{\mu}$. This precisely gives $\Delta s=\frac{\pi \mu}{2 p_{\psi}}$.}. Hence $(\tilde{\mathcal{M}},\tilde{g})$ is not a causal geodesically complete space-time. In the case where $c\ne0$ and $p_\psi\ne0$ the turning point will depend on the ratio $c/p_\psi^2$, which we can adjust either if we want a turning point in the region $\{\frac{1}{\mu}< y(s)< y^{-}\}$ or if we do not. Thus, not all null geodesics with $p_\phi=0$ will reach the singularity.

Now, we consider geodesic segments for which $y(s)\ge\frac{1}{\mu}+\varepsilon$, with any $\varepsilon>0$. In this case, looking at Eq.\eqref{tc}, there exists a constant $C_{12}$ such that
\begin{equation} \label{c5}
\frac{R^2F^2(y)F(x)}{(x-y)^2}\left(\frac{\dot{y}^2}{|G(y)|}+\frac{\dot{x}^2}{G(x)}\right)+\frac{(x-y)^2F^2(x)}{R^2G(x)F^2(y)}p_{\phi}^2\le C_{12}.
\end{equation}
It follows from here that a non-zero $p_{\phi}$ prevents $x$ from approaching $\pm1$, thus, since that the coefficients of $\dot{y}^2$ and $\dot{x}^2$ are bounded from above and away from zero we have that
\begin{equation} \label{c6}
\dot{y}^2+\dot{x}^2\le C_{13}.
\end{equation}
Also in the case $p_{\phi}=0$, from Eq.\eqref{c5} we find bounds on $|\dot{y}|,|\dot{x}|$ obtaining again Eq.\eqref{c6}. Moreover, in the case $p_{\phi}\ne0$ from \eqref{cphi} we find a bound on $|\dot{\phi}|$ and from \eqref{c5}, with $p_\phi$ as in \eqref{cphi}, otherwise. Also, from \eqref{cpsi} and \eqref{ttt} we find immediately bounds on $|\dot{\psi}|$ and $|\dot{t}|$ respectively. Thus, in general we have
\begin{equation}
\dot{t}^2+\dot{y}^2+\dot{x}^2+\dot{\phi}^2+\dot{\psi}^2\le C_{14}.
\end{equation}
This, together with the bounds on $\dot{y}$ and $\dot{\theta}$ towards $y\to y^-$ which will be derived in the next section, concludes the proof of Prop.\ref{ptc}. 

We would like to end this section by noting that a similar lemma to \ref{lem1} can be written down for the neutral case. In order to do so note that in the neutral limit $\mu=\lambda$ and $\hat{G}(\xi)$ reduces to \eqref{ghat}, thus we are working in the region $\{y^{-}< y(s)< y^+\}$ as above but now with $y^-$ representing the curvature singularity. So from \eqref{eqlo1}, looking at the leading order terms close to $y=\frac{1}{\lambda}$, and from Eqs.\eqref{eqh2},\eqref{eqh3} we obtain
\begin{equation}\label{boun1}
\dot{y}\le \frac{\varepsilon}{\sqrt{|F(y)|}},
\end{equation}
for some $\varepsilon>0$, which is valid for both timelike and null geodesics. Rewriting Eq.\eqref{lagin} in the neutral limit leads to
\begin{equation} \label{eqnl}
\frac{|F(y)|}{|G(y)|}\dot{y}^2-\frac{F^2(y)}{F(x)}\dot{\theta}^2-\frac{(x-y)^4}{R^2G(x)F^2(y)}p_\phi^2=\Theta,
\end{equation}
with
\begin{equation}
\Theta=\frac{(x-y)^2}{R^2F(x)}\left(-\mathcal{L}+\frac{|F(y)|}{F(x)}p_{t}^2\right)+\frac{(x-y)^4}{R^2F^2(x)|G(y)|}(RC(1+y)p_{t}-p_{\psi})^2.
\end{equation}
Now, the first term on the LHS of \eqref{eqnl} is bounded as $y\to\frac{1}{\lambda}$ due to \eqref{boun1} and moreover the RHS is bounded and positive, thus, since the second term on the LHS is negative and the third necessarily diverges as $y\to\frac{1}{\lambda}$ we must have $p_\phi=0$ for causal geodesics reaching the singularity. Thus we can write down the lemma:
\begin{lemma}
In the neutral singly spinning black ring space-time, causal geodesics reaching the singular boundary $\{y=\frac{1}{\lambda}\}$ must have $p_\phi=0$.
\end{lemma}
This is similar to the geodesic structure of the 4-dimensional Schwarzschild black hole, where causal geodesics that hit the singularity must have zero angular momentum (see \cite{Graves:1960zz}). 

\subsubsection*{Geodesics in the regions $\{y^+< y(s)\le2 y^{+}\}$ and $\{\frac{y^-}{2}\le y(s)<y^{-}\}$}
In these regions both $G(y)$ and $F(y)$ are negative and thus we rewrite \eqref{lag} as
\begin{equation} \label{eqpm}
\begin{split}
F^2(y)\left(\frac{\dot{y}^2}{|G(y)|}+\frac{\dot{\theta}^2}{F(x)}\right)&+\frac{(x-y)^4F(x)}{R^4G(x)F^2(y)}p_{\phi}^2+\frac{(x-y)^2}{R^2F(x)}\frac{|F(y)|}{F(x)}p_{t}\\
&-\frac{(x-y)^4|F(y)|}{R^4F^2(x)|G(y)|}(RC(1+y)p_{t}-p_{\psi})^2=\frac{(x-y)^2}{R^2F(x)}\mathcal{L},
\end{split}
\end{equation}
and we wish to show that
\begin{proposition} \label{pjo}
All maximally extended causal geodesics through $\mathcal{M}_{I},\mathcal{M}_{V}$, are either complete (or reach the singular boundary $\{y=\frac{1}{\mu}\}$ in finite affine time) or can be smoothly extended across the Killing horizons $\{y=y^\pm\}$.
\end{proposition}
Note that geodesics which stay bounded away from $y^\pm$ have been considered in the previous analysis so assume that there exists a sequence of $s_{i}\to\hat{s}$ such that $y(s_{i})\to y^\pm$. Now, to show that the above statement is true first note that $\xi^{\pm}$, given in \eqref{av}, are the Killing vector fields\footnote{We can write \eqref{av} in coordinates $(t,y,\psi,x,\phi)$ by just replacing $v$ with $t$ and $\hat{\psi}$ with $\psi$ since the transformation \eqref{cin} preserves the Killing vectors.} tangent to the horizons $y=y^\pm$, and since the horizons are non-degenerate, $\xi^{\pm}$ are timelike for a small enough region close to $y=y^\pm$. Hence we must have for any future directed causal geodesic that $g(\xi^\pm,\dot{\gamma})<0$, which implies
\begin{equation} \label{cond1}
RC(1+y^\pm)p_{t}-p_{\psi}\ne0.
\end{equation} 
Alternatively, suppose that at $y=y^\pm$, $RC(1+y^\pm)p_{t}-p_{\psi}=0$. In this case we have that 
\begin{equation} \label{cond11}
RC(1+y^\pm)p_{t}-p_{\psi}=RC(y-y^\pm)p_t.
\end{equation}
Inserting this into Eq.\eqref{eqpm} implies that as $y\to y^\pm$ the last term on the LHS vanishes. Noting that the first four terms on the LHS are positive and in the case of timelike geodesics the RHS is negative, then such geodesics cannot reach the horizon. Moreover if null geodesics are to reach the horizon they must satisfy $\dot{y}=\dot{\theta}=p_\phi=p_t=0$ at $y=y^\pm$. But then $p_t=0$ implies that the last term on LHS vanishes always due to \eqref{cond11} and thus that $\dot{y}=\dot{\theta}=0$ always. Therefore we conclude again \eqref{cond1}.

Now, the analysis for both regions $\{y^+< y(s)\le2 y^{+}\}$ and $\{\frac{y^-}{2}\le y(s)< y^{-}\}$ is essentially the same so we focus on the region $\{y^+< y(s)\le2 y^{+}\}$. We continue by noting that Eq.\eqref{eqh} holds for any interval of $s$ for which $\dot{y}$ does not change sign. We now want to show that for future directed causal geodesics sufficiently close to the horizon $\dot{y}$ is negative. So suppose not, that there will be increasing sequences $\{ s_{i}^\pm\}_{i\in N}$, $s_{i}^\pm\to \hat{s}$, with $s_{i}^-<s_{i}^+$ such that
\begin{equation} \label{contra}
\dot{y}(s_{i}^{\pm})=0, ~\dot{y}<0~\text{on}~I_{i}:=(s_{i}^-,s_{i}^+),~y_{i}^\pm:=y(s_{i}^\pm)\searrow y^+,~y_{i}^->y_{i}^+ .
\end{equation}
Then from Eq.\eqref{eqlo1} and Eq.\eqref{eqh2} there exists a $y_{*}>y^+$ and $\varepsilon>0$ such that for all $y\in(y^+,y_{*})$ we have
\begin{equation}
\frac{d\hat{h}^2}{dy}\le-\frac{\varepsilon}{(y-y^+)^2}.
\end{equation}
Integrating this gives
\begin{equation}
\hat{h}^2(y_{i}^-)-\hat{h}^2(y_{i}^+)\le-\int_{y_{i}^+}^{y_{i}^-}\frac{\varepsilon}{(y-y^+)^2}dy<0,
\end{equation}
which contradicts the original assumption that $\dot{y}(s_{i}^\pm)=0$ and hence $\dot{y}$ is strictly negative sufficiently close to the event horizon. We note that if we had considered the region $\{\frac{y^-}{2}< y(s)\le y^{-}\}$ we would had found $\dot{y}$ to be strictly positive.

Now, as in the region $\{y^{-}< y(s)< y^+\}$ we find Eq.\eqref{eqy1}, i.e., $|\dot{y}|$ is bounded away from zero. Moreover, from \eqref{eqpm} multiplied by $(ds/dy)^2$ we find the bound
\begin{equation} \label{tz12}
\left|\frac{d\theta}{dy}\right|\le\frac{C_{15}}{\sqrt{|G(y)|}}.
\end{equation}
We can now repeat the arguments starting around Eq.\eqref{eqy1} but with Eq.\eqref{tz2} replaced by Eq.\eqref{tz12} and the conclusion of  Prop.\ref{pjo} follows. Furthermore, collecting the results from the previous sections we are lead to the following theorem:
\begin{theorem} \label{teotim}
All maximally extended timelike geodesics in $(\tilde{\mathcal{M}},\tilde{g})$ are complete, never reaching the singular boundary $\{y=\frac{1}{\mu}\}$.
\end{theorem}
This resembles the geodesic structure of the 4-dimensional RN black hole (see \cite{Graves:1960zz}), where we also have timelike completeness.

\subsubsection*{Geodesics at the Killing horizons $y=y^\pm$}
In terms of the coordinates $(\hat{v}^\pm,\hat{u}^\pm,\hat{\psi}^\pm)$ the Killing horizons sit at $\{\hat{v}^{\pm}\hat{u}^\pm=0\}$. Hence if a timelike geodesic  $\gamma(s)$ satisfies $\gamma(s)\in\{\hat{v}^{\pm}\hat{u}^\pm=0\}$ for some $s$ and if $\frac{d\hat{v}^{\pm}}{ds}$ and $\frac{d\hat{u}^{\pm}}{ds}$ are different than zero then the geodesic $\gamma(s)$ will enter one of the regions analyzed above and hence is complete, otherwise, in the case where $\frac{d\hat{v}^{\pm}}{ds}$ or $\frac{d\hat{u}^{\pm}}{ds}=0$, the geodesics must be lightlike since these geodesics correspond to the null generators of the bifurcate Killing horizons $\mathcal{H}^\pm$, which have been shown to be complete \cite{Boyer:1969c}. Now, collecting the results in all regions completes the proof of Theo.\ref{tmax} and hence of Theo.\ref{tmaxi} for the non-extremal case.

\section{Extensions for extremal dipole black rings} \label{sex}
In the extremal case $\lambda=\nu\equiv y_h^{-1}$ and the function $G(y)$ acquires a second order zero at $y=y_h$. Global coordinates are more difficult to find, nevertheless, we can construct different Eddington-Finkelstein-type extensions for extremal dipole black rings. In this section, this task will be carried out using the methods developed in \cite{Durkee:2008an},\cite{Kunduri:2007vf},\cite{Chrusciel:2009vr}.  Later, using one of these extensions we introduce a set of ingoing  $(v,\hat{\psi})$ and outgoing $(u,\hat{\psi})$ coordinates, valid inside the ergoregion and the horizon, and use the coordinate system $(t,y)$ introduced in \eqref{ds1}, also valid outside the ergosurface, to construct the full causal diagram and analyze some of its properties.

\subsection{Eddington-Finkelstein-type coordinates} \label{inout}
Here we construct ingoing and outgoing coordinates for the dipole black ring using two different methods. First, we find a solution for the Hamilton-Jacobi equation representing zero energy null geodesics and then use these to construct a new coordinate system valid for both extremal and non-extremal dipole black rings. This approach was first done in \cite{Durkee:2008an} for doubly-spinning black rings. Second, we construct a different set of coordinates for extremal dipole black rings using a method which was originally developed for taking the near-horizon limit of extremal black hole solutions \cite{Kunduri:2007vf} but later shown that it could also be used for the complete solution of the doubly-spinning black ring \cite{Chrusciel:2009vr}. 

\subsubsection{Following zero energy null geodesics} \label{f0z}
The starting point of the construction is to consider the Hamiltonian for a particle moving in the background $g$, which can be constructed from the Lagrangean \eqref{lag2} in the following way
\begin{equation} \label{ham}
\mathcal{H}(x^{\mu},p_{\nu})=p_{\mu}\dot{\gamma}^{\mu}-\mathcal{L}(x^{\mu},\dot{\gamma}^{\nu})=\frac{1}{2}g^{\mu\nu}p_{\mu}p_\nu ,
\end{equation}
where $p_{\nu}$ is the conjugate momenta derived from \eqref{lag2}, and its associated Hamilton-Jacobi (HJ) equation
\begin{equation} \label{hj1}
\frac{\partial S}{\partial s}+\mathcal{H}\left(x^{\mu},\frac{\partial S}{\partial x^{\nu}}\right)=0.
\end{equation}
Now, in order to find separable solutions to Eq.\eqref{hj1} we first need the conjugate momenta $p_\mu$ which can be easily obtained from the Lagrangean \eqref{lag2} and read
\begin{equation} \label{p1}
E\equiv -p_{t}=\frac{F(x)}{F(y)}(\dot{t}+RC(1+y)\dot{\psi}),~p_{\psi}=-RC(1+y)E-\frac{R^2F(x)G(y)}{(x-y)^2F(y)}\dot{\psi},
\end{equation}
\begin{equation} \label{p2}
p_{\phi}=\frac{R^2F(y)^2G(x)}{F(x)^2}\dot{\phi},~p_{x}=\frac{R^2F(y)^2F(x)}{(x-y)^2G(x)}\dot{x},~p_{y}=-\frac{R^2F(y)^2F(x)}{(x-y)^2G(y)}\dot{y}.
\end{equation}
As noted in Sec.\ref{snex}, the three constants $p_{t},p_\psi ,p_\phi$ are the conserved charges associated with the Killing vector fields $\partial_t , \partial_\psi , \partial_\phi$ and hence conserved along any geodesic. The Hamiltonian \eqref{ham} is a 5-dimensional system so we need 5 constants of motion in order to have complete integrability of the system. The fourth one can be acquired by imposing the mass shell condition $g^{\mu\nu}p_{\mu}p_{\nu}=-m^2$, while the remaining one can in principle be obtained from the HJ equation. 

From what has been said, we now make an ansatz
\begin{equation}
S(s,t,y,\psi,x,\phi)=\frac{1}{2}m^2 s-Et+p_\phi \phi + p_\psi \psi + S_{x}(x)+S_{y}(y),
\end{equation}
where $S_{x}(x)$ and $S_{y}(y)$ are arbitrary functions of $x$ and $y$ respectively. This ansatz is the same as the one taken in \cite{Durkee:2008an} for doubly spinning black rings and should in principle also leave the HJ equation into a separable form due to the similarity between both metrics. 

Inserting the ansatz into Eq.\eqref{hj1} we obtain
\begin{equation}
\begin{split}
G(y)\left(\frac{dS_{y}}{dy}\right)^2 -& G(x)\left(\frac{dS_{x}}{dx}\right)^2=R^2F(x)F(y)^2m^2+\frac{F(x)^3}{G(x)}p_{\phi}^2-\frac{F(y)^3}{G(y)}p_{\psi}^2\\
&-\frac{2RC(1+y)F(y)^3}{G(y)}Ep_\psi-\left(\frac{R^2C^2F(y)^3(1+y)^2}{G(y)}+\frac{R^2F(y)^3}{(x-y)^2}\right)E .
\end{split}
\end{equation}
After a short inspection of this equation one can easily realize that it is not possible in general to separate the $m^2$ and $E$ terms. So, to get out some form of separability we restrict ourselves to zero energy ($E=0$) null geodesics ($m=0$) leading to
\begin{equation} \label{hj2}
G(y)\left(\frac{dS_{y}}{dy}\right)^2+\frac{F(y)^3}{G(y)}p_{\psi}^2=G(x)\left(\frac{dS_{x}}{dx}\right)^2+\frac{F(x)^3}{G(x)}p_{\phi}^2=c,
\end{equation}
where $c$ is a constant, which has the physical interpretation of being the extra conserved quantity associated with these geodesics, and describes all possible zero energy null geodesics.

Now, we want to show that these geodesics go through the horizon and then use them to construct extensions valid through the future and past event horizons. So, plugging in $p_{x}$ and $p_{y}$ given in \eqref{p2} into Eq.\eqref{hj2} we obtain the following two equations governing the geodesics along the $x$ and $y$ directions:
\begin{equation} \label{hj3}
\dot{x}^2+\frac{(x-y)^4}{R^4F(y)^4F(x)^2}\left[F(x)^3p_\phi^2-G(x)c\right]=0,~\dot{y}^2+\frac{(x-y)^4}{R^4F(y)^4F(x)^2}\left[F(y)^3p_\psi^2-G(y)c\right]=0.
\end{equation}
The analysis of these geodesics is essentially the same as for the neutral singly spinning case studied in \cite{Durkee:2008an}. We have three cases to consider: \\ \\
\textbf{Case 1:} $c=0$. \\ First note that, looking at \eqref{hj3}, since $F(x)$ and $G(x)$ are always positive we must have that $c\ge0$ since otherwise we could not have any solution for $\dot{x}$. This implies, from \eqref{hj3}, that these geodesics are only physically realizable in the ergoregion. Now, taking $c=0$, since $F(x)>0$ we must have $p_\phi=0$ in order to have a solution for $\dot{x}$. Also, since $F(y)<0$ we must have $p_\psi\ne0$ if we want to have a path at all. Therefore, from \eqref{hj3} we obtain
\begin{equation} \label{hj4}
\dot{x}^2=0,~\dot{y}^2+\frac{(x-y)^4}{R^4F(y)F(x)^2}p_\psi^2=0.
\end{equation}
Also, since $F(y)<0$ in the ergoregion, the only turning point $\dot{y}=0$ is at the ergosurface where $F(y)\to+\infty$\footnote{In order to see this explicitly one has to change to $Y$ coordinates.}. Thus such paths represent geodesics that came from the past event horizon turned around at the ergosurface and eventually crossed the future event horizon. \\ \\
\textbf{Case 2:} $c>0$ and $p_\phi=0$. \\ In this case, since $G(y)<0$ in the ergoregion, the turning point $\dot{y}=0$ moves inwards towards the horizon as the ratio $c/p_\psi^2$ is increased. Also since $cG(x)>0$ everywhere $x$ can take in value on the interval $-1\le x\le1$. Thus these solutions correspond to geodesics that rotate around the $S^2$ as they fall into the horizon.\\ \\
\textbf{Case 3:} c>0 and $p_\phi>0$. \\ In this case, the behavior in the $y$ direction remains qualitatively the same but the motion in the $x$ direction changes. It is easy to check that since $p_\phi\ne0$ and $F(x)>0$ the range of $x$ is shortened as $p_\phi^2/c$ is increased. These geodesic fall into the horizon but their movement in the $S^2$ is restricted to a shorter $x$ interval. \\

Knowing that these geodesics do go through the horizon, we now want to use them to construct ingoing and outgoing null coordinates regular at the horizon. In order to accomplish this task note first that zero energy null geodesics must satisfy the equations of motion given by Eqs.\eqref{p1} and \eqref{p2} and \eqref{hj3} and read
\begin{equation} \label{path}
\begin{split}
\dot{t}=\frac{C(1+y)F(y)(x-y)^2}{RF(x)G(y)}&p_\psi,~\dot{\psi}=-\frac{F(y)(x-y)^2}{R^2F(x)G(y)}p_\psi,~\dot{\phi}=\frac{F(x)^2(x-y)^2}{R^2F(y)^2G(x)}p_\phi, \\
\dot{y}=\frac{(x-y)^2}{R^2F(x)F(y)^2}&\sqrt{\zeta_{y}(y)},~\dot{x}=\pm\frac{(x-y)^2}{R^2F(y)^2F(x)}\sqrt{\zeta_{x}(x)},
\end{split}
\end{equation}
where we have defined the functions
\begin{equation}
\zeta_{y}(y)=-F(y)^3p_\psi^2+G(y)c,~\zeta_{x}(x)=-F(x)^3p_\phi^2+G(x)c .
\end{equation}
Note that we have taken the $-$ sign in the expression for $\dot{y}$ in order to obtain ingoing coordinates, outgoing coordinates can be obtained by just flipping the sign in the end. We want to find a set of coordinates which are constant along these geodesics, moreover, if we want to preserve the $x,y$ symmetry of the original metric \eqref{ds1} we must find functions $\eta^{i}(x,y)$ for $i=t,\psi,\phi$ such that
\begin{equation} \label{path1}
\dot{t}-\frac{\partial \eta^t}{\partial x}\dot{x}-\frac{\partial \eta^t}{\partial y}\dot{y}=\dot{\psi}-\frac{\partial \eta^\psi}{\partial x}\dot{x}-\frac{\partial \eta^\psi}{\partial y}\dot{y}=\dot{\phi}-\frac{\partial \eta^\phi}{\partial x}\dot{x}-\frac{\partial \eta^\phi}{\partial y}\dot{y}=0,
\end{equation}
and hence the coordinates $v=t-\eta^t$, $\hat{\psi}=\psi-\eta^\psi$ and $\hat{\phi}=\phi-\eta^{\phi}$ will be constant along these geodesics. Inserting \eqref{path} into \eqref{path1} it is straightforward to find that the functions $\eta^i(x,y)$ must satisfy
\begin{equation}
\begin{split}
\eta^t=-RCp_\psi  \int_{y_0}^{y} &\frac{(1+y')F(y')^3}{G(y')\sqrt{\zeta_{y}(y')}}dy',~\eta^\psi=p_\psi\int_{y_0}^{y}\frac{F(y')^3}{G(y')\sqrt{\zeta_{y}(y')}}dy',\\
\eta^\phi=&\pm p_\phi\int_{x_0}^{x}\frac{F(x')^3}{G(x')\sqrt{\zeta_{x}(x')}}dx',
\end{split}
\end{equation} 
where the choice $x_{0}=0$ and $y_0$ being the turning point of geodesic assures a well defined integral. The resulting coordinate differentials are thus:
\begin{equation} \label{nc}
dv=dt+RCp_\psi\frac{(1+y)F(y)^3}{G(y)\sqrt{\zeta_{y}(y)}}dy,~d\hat{\psi}=d\psi-\frac{p_\psi F(y)^3}{G(y)\sqrt{\zeta_{y}(y)}}dy,~d\hat{\phi}=d\phi\mp \frac{p_\phi F(x)^3}{G(x)\sqrt{\zeta_{x}(x)}}dx.
\end{equation}
Performing this coordinate transformation on the original metric \eqref{ds1} yields a new metric in coordinates $(v,y,\hat{\psi},x,\hat{\phi})$ which reads:
\begin{equation} \label{ds5}
\begin{split}
&ds^2=-\frac{F(x)}{F(y)}\left(dv+RC(1+y)d\hat{\psi}\right)^2 \\
&+\frac{R^2F(y)^2F(x)}{(x-y)^2}\left[\frac{cdx^2}{\zeta_x(x)}-\frac{cdy^2}{\zeta_y(y)} +\frac{G(x)}{F(x)^3}d\hat{\phi}^2-\frac{G(y)}{F(y)^3}d\hat{\psi}^2\pm\frac{2p_\phi}{\sqrt{\zeta_x(x)}}d\hat{\phi}dx-\frac{2p_\psi}{\sqrt{\zeta_y(y)}}d\hat{\psi}dy  \right].
\end{split}
\end{equation}
Since the Jacobian of the transformation \eqref{nc} is equal to one and the coefficients of this new metric are analytic at $y=y_h$ then so are the coefficients of the inverse metric and hence we can extend the space-time smoothly across the Killing horizons. Also, in the simplest case where $p_\phi=0$, $c=0$ and $p_\psi>0$ we obtain the coordinate differentials \eqref{cin} and the metric reduces to
\begin{equation} \label{ds2}
\begin{split}
ds^2=&-\frac{F(x)}{F(y)}\left(dv+RC(1+y)d\hat{\psi}\right)^2 \\
&+\frac{R^2}{(x-y)^2}\left[F(x)\left(-\frac{G(y)}{F(y)}d\hat{\psi}^2+2\sqrt{-F(y)}d\hat{\psi}dy\right)+F(y)^2\left(\frac{F(x)}{G(x)}dx^2+\frac{G(x)}{F(x)^2}d\phi^2\right)\right].
\end{split}
\end{equation}
As a final comment we note that these coordinates are not of the Eddington-Finkelstein-type in the usual sense since they are only valid in the ergoregion (and also inside the horizon) where $F(y)<0$, which is clear from the square root term in $g_{v\hat{\psi}}$ in \eqref{ds2}. Below we construct coordinates of the same type which cover the full original space-time given in \eqref{ds1}.

\subsubsection{Near-horizon limit-type extension}

As in \cite{Kunduri:2007vf},\cite{Chrusciel:2009vr} we must start with a metric in the form
\begin{equation} \label{ds3}
ds^2=g_{tt}dt^2+2g_{ti}dtd\Phi^{i}+g_{yy}dy^2+g_{xx}dx^2+g_{ij}d\Phi^{i}d\Phi^{j},
\end{equation}
where the functions $g_{tt},g_{ti}$ should vanish at the horizon $y_{h}$ that we want to extend through and have the following behavior
\begin{equation}
g_{tt}=f_{t}(x,y)(y-y_{h})^2,~g_{ti}=f_{i}(x,y)(y-y_{h}),~g_{yy}=\frac{h(x,y)}{(y-y_{h})^{2}},
\end{equation}
where $f_{t},f_{i},h$ are well behaved functions and non-zero at $y=y_{h}$. The dipole black ring metric \eqref{ds1} does not satisfy the requirement that $g_{tt},g_{ti}$ should vanish at the horizon with second order and first order zeros respectively, so we begin by performing the transformation
\begin{equation}
d\hat{\psi}=d\psi-adt,
\end{equation} 
where $a$ is a constant to be adjusted. The requirement of such behavior implies again that $a$ should be given as in \eqref{crot} and hence the metric takes the form of \eqref{ds2}, where $\Phi^{i}=(\psi,\phi)$. The metric is still singular at $y=y_{h}$, so in order to remove the singularity at the horizon we introduce a new set of coordinates $(v,z,\varphi)$ such that
\begin{equation}
dt=dv+a(y)dy,~d\varphi=d\hat{\psi}+b(y)dy,~dy=dz,
\end{equation}
where $a(y)$ and $b(y)$ are functions to be chosen shortly. The Jacobian of the transformation is simply equal to one so we just need to check that the metric functions are well behaved. As in \cite{Kunduri:2007vf},\cite{Chrusciel:2009vr} we now make the choice
\begin{equation}
a(y)=\frac{a_{0}}{(y-y_{h})^2}+\frac{a_{1}}{y-y_h},~ b(y)=\frac{b_0}{y-y_h}.
\end{equation}
The metric coefficients then take the form
\begin{equation}
\begin{split}
g_{vz}=&(a_{1}(y-yh)+a_0)f_{t}+b_0f_\psi, \\
g_{\varphi z}=& \left(a_{1}f_\psi (y-y_h) + a_0 f_\psi + b_0 g_{\hat{\psi}\hat{\psi}}\right)\frac{1}{y-y_h}, \\
g_{zz}=&\left((a_1(y-y_h)+a_0)^2f_t + 2f_\psi (a_1(y-y_h) +a_0) b_0 + b_{0}^2  g_{\hat{\psi}\hat{\psi}} + h\right)\frac{1}{(y-y_h)^2},\\
g_{vv}=&f_{t}(y-y_h)^2,~g_{v\varphi}=f_\psi (y-y_h).
\end{split}
\end{equation}
The metric functions $g_{vv},g_{v\varphi}$ and $g_{vy}$ are well behaved,  hence, in order to obtain a smooth metric at $y=y_{h}$ we must require
\begin{equation}
\begin{split}
(y-y_h)g_{z\varphi}|_{y=y_{h}}=0,~(y-y_h)^2g_{zz}|_{y=y_{h}}=0,~\partial_{z}((y-y_h)^2g_{zz})|_{y=y_h}=0,
\end{split}
\end{equation}
leading to the following system of equations
\begin{equation}
\begin{split}
a_{0}f_\psi+b_0 g_{\hat{\psi}\hat{\psi}}=&0\\
a_{0}^2f_t+2a_0 b_0 f_\psi+b_{0}^2 g_{\hat{\psi}\hat{\psi}} + h =&0\\ 
2a_0 a_1 f_t +a_{0}^2\partial_z f_{t} + 2a_1 b_0 f_\psi + 2a_0 b_ 0\partial_z f_\psi + b_{0}^2\partial_z  g_{\hat{\psi}\hat{\psi}} + \partial_z h=&0
\end{split}
\end{equation}
where
\begin{equation}
\begin{split}
f_{t}(x,y)=-& \frac{F(x)}{F(y)(1-y_h)^2}-\frac{F(x)(1-y^2)}{R^2(1+y_h)^2(x-y)^2F(y)}, \\
f_\psi(x,y)=\frac{RF(x)(1+y)}{y_h(1+y_h)F(y)}&+\frac{F(x)(1-y^2)(y-y_h)}{RC(1+y_h^2)^2(x-y)^2F(y)},~ h(x,y)=-\frac{R^2y_{h}^2F(x)F(y)^2}{(x-y)^2(1-y^2)},
\end{split}
\end{equation}
with $g_{\hat{\psi}\hat{\psi}}=g_{\psi\psi}$ as given by the coordinates in \eqref{ds1}. After cumbersome calculations one can find that the above system of equations can be solved by choosing one of the two possible triads
\begin{equation}
(a_0 , a_1 , b_0)=\left(\pm \frac{Ry_h(2+(y_h-3)\mu)\sqrt{y_h\mu-1}}{2(1-y_h)^2} ,\pm\frac{Ry_h(\mu y_h - 1)^{\frac{3}{2}}}{y_h -1},\pm \frac{y_h^2 (\mu y_h-1)^{\frac{3}{2}}}{(y_h -1)(1+y_h)^2} \right),
\end{equation}
where the $+$ sign corresponds to an extension through the future event horizon and the $-$ sign to an extension through the past event horizon.

\subsection{Causal structure}
\begin{wrapfigure}{r}{0.3\textwidth} 
\vspace{-35pt}
  \begin{center}
    \includegraphics[width=0.25\textwidth, height=0.4\textheight]{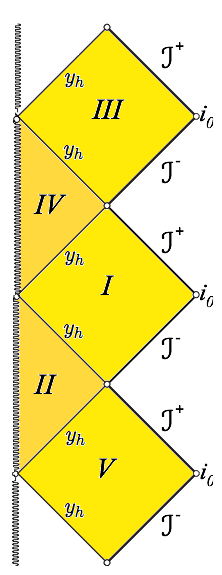}\label{fig:ex}
  \end{center}
  \vspace{-20pt}
  \caption{\small Causal structure of the extremal dipole black ring generated via the iteration procedure explained on the left.}
  \vspace{-40pt}
  \label{fig2}
\end{wrapfigure}
The causal structure of the extremal dipole black ring is depicted in Fig.\ref{fig2} and resembles the four dimensional extremal Reissner-Nordstr\"{o}m black hole. We obtained this diagram using some of the results in the sections below and by the following iterative procedure. First define the original space time in the coordinates of \eqref{ds1} as $\mathcal{M}_{I}$, which covers the interval $\{y_{h}<y\le-1\}$. Next,  introduce ingoing coordinates $(v_{I},\hat{\psi}_{I})$ in this region as in \eqref{cin} with $\lambda=\nu=y_h^{-1}$, where the index $I$ denotes the region $I$ where $(v,\hat{\psi})$ have been introduced. The metric becomes analytic at the future event horizon $y=y_{h}$ and hence also analytic in $\mathcal{M}_{IV}$ which covers $\{\frac{1}{\mu}<y<y_{h}\}$, with the form \eqref{ds2}. The coordinate $v_{I}$ belongs to the interval $(-\infty,\infty)$ but can be extended through $\{v_{I}=\pm\infty\}$ by changing to outgoing coordinates described below. Similarly, instead of introducing initially the coordinates $(v_{I},\hat{\psi}_{I})$, we could have introduced outgoing coordinates $(u_{I},\hat{\psi}_{I})$ of the form
\begin{equation} \label{cout}
\begin{split}
du=&dt+RC(1+y)\frac{F(y)\sqrt{-F(y)}}{G(y)}dy,\\
d\hat{\psi}=&d\psi-\frac{F(y)\sqrt{-F(y)}}{G(y)}dy,
\end{split}
\end{equation}
and analytically continued through the past event horizon to the region denoted by $\mathcal{M}_{II}$. The metric \eqref{ds2} takes the same form but the component $g_{y\hat{\psi}}$ acquires an overall minus sign. In this new region where $\{\frac{1}{\mu}<y<y_{h}\}$ we can transform back to coordinates $(t,\psi)$ given in \eqref{ds1} by using the inverse transformation which can be obtained from \eqref{cout}. If we now introduce again $(v_{II},\hat{\psi}_{II})$ coordinates we see that the surfaces $\{v_{II}=+\infty\}$ and $\{v_{I}=-\infty\}$ are identified. Moreover, in the region $M_{IV}$ we can introduce $(t,\psi)$ coordinates and then outgoing coordinates $(u_{IV},\hat{\psi}_{IV})$ which allows us to cross the surface $\{v_{II}=+\infty\}$ to a new asymptotic flat region denoted by $\mathcal{M}_{III}$. If we define $\mathcal{M}_{1}=\mathcal{M}_{I}\cup\mathcal{M}_{II}\cup\mathcal{M}_{III}\cup\mathcal{M}_{IV}$, we can then construct the remaining diagram by patching together infinite copies of $\mathcal{M}_{1}$ labeled as $\mathcal{M}_{n}$. We name the space-time so constructed as $(\tilde{\mathcal{M}},\tilde{g})$.

\subsection{Black hole and white hole regions}
To show that the sets $B:=\{\mathcal{M}_{IV}\}$ and $W:=\{\mathcal{M}_{II}\}$ are black hole and white hole regions with respect to an observer in $\mathcal{M}_{I}$ we cannot use the same arguments as in the discussion around \eqref{hyp1} because since now $G(y)$ has a second order zero at $y=y_{h}$ it does not change sign when crossing to the region $y<y_{h}$. Therefore we proceed differently. We introduce yet another set of coordinates, which will be useful for the purpose, which can be obtained by combining the $(v,u)$ coordinates defined in both transformations \eqref{cin} and \eqref{cout}. The metric coefficients take the same form as in \eqref{m1} but now with $a=0$ and
\begin{equation}
H(y)=-\frac{G(y)}{RC(1+y)F(y)\sqrt{-F(y)}}.
\end{equation}
Inspecting \eqref{m1} we see that the metric functions are well behaved in the interval $\{y_{h}<y<-\infty\}$ and inclusively at $y=y_{h}$ but the determinant is singular there, as in most of the double null forms of the metrics known in the literature \cite{Townsend:1997ku},\cite{Hawking:ls}. 

Now, by the method explained above we can analytically continue from $\mathcal{M}_{I}$ to $\mathcal{M}_{IV}$, where we can change back to $(t,\psi)$ coordinates and then introduce $(v_{IV},u_{IV})$ coordinates in the way defined above\footnote{We will now drop the index $IV$ for simplicity.}. In terms of these coordinates we note that the coordinate differential $dt$ is given by \eqref{diff1}, thus we have that 
\begin{equation}
\dot{t}=\frac{1}{2}(\dot{v}+\dot{u}).
\end{equation}
Moreover, for future directed causal geodesics in the region $\mathcal{M}_{IV}$ we must have $\dot{t}>0$ since $t$ is a time function there, and in fact in every region of the space-time since G(y) never changes sign, and thus $\dot{v}+\dot{u}>0$. Also, integrating Eqs. \eqref{cin} and \eqref{cout} we obtain
\begin{equation} \label{hc}
v=t+K(y),~u=t-K(y),
\end{equation}
where $K(y)$ is a complicated function of $y$ that as $y\to y_{h}$, $K(y)\to-\infty$ if we approach $y_{h}$ from above and $K(y)\to+\infty$ if otherwise. From \eqref{hc} and from the behavior of $K(y)$ as $y\to y_{h}$ we see that the horizon that separates the regions $\mathcal{M}_{IV}$ from $\mathcal{M}_{I}$ is reached when $u\to-\infty$. We wish now to show that future directed causal geodesics in $\mathcal{M}_{IV}$ cannot cross back to $\mathcal{M}_{I}$ through the surface $\{u=-\infty\}$. From \eqref{cout} we can also obtain
\begin{equation}\label{hc1}
\dot{u}=\dot{t}+RC(1+y)\frac{F(y)\sqrt{-F(y)}}{G(y)}\dot{y}.
\end{equation}
Therefore, for future directed causal geodesics approaching $y_{h}$ from below we must have that at least close to $y_{h}$, $\dot{y}>0$ if the geodesic is to cross the horizon, in fact, we show this explicitly in the section below. This implies that $\dot{u}>0$ for future directed causal geodesics crossing the horizon, and that from \eqref{hc} $v\to+\infty$. Hence, as $t$ increases, these geodesics get further away from the surface $\{u=-\infty\}$ and are forced to cross the surface $\{v=+\infty\}$. Thus they cannot cross back to $\mathcal{M}_{I}$. Similar arguments apply to $\mathcal{M}_{II}$ and thus $B$ and $W$ are indeed black hole and white hole regions with respect to an observer in $\mathcal{M}_{I}$, as long as the Killing horizon coincides with the event horizon, which we will show below to be the case. A more general argument which applies to this case also is given in \cite{Chrusciel:2010ix} (section 5.7).

\subsection{Horizon topology, maximality and non-global hyperbolicity}
The arguments for this section follow closely the ones in Sec.\ref{snex}. We start by noting that since $G(y)$ is always negative there must exist always a timelike combination of Killing vector fields in the \emph{d.o.c.}, and thus the event horizon must coincide with the Killing horizon with topology $S^{1}\times S^{2}$. 

Regarding maximality, we can show that a part of the geodesic analysis done in the non-extremal case in Sec.\eqref{snex} holds here, more specifically, Prop.\ref{pinout} holds in the region $\{y_h<y<+\infty\}\cup\{-\infty<y\le-1\}$ and Prop.\ref{ptc} holds in the region $\{\frac{1}{\mu}\le y(s)\le(y_{h}-\epsilon)\}$. However, extendibility of geodesics in the regions $\{y_{h}<y(s)\le2y_{h}\}$ and $\{\frac{y_{h}}{2}\le y(s)<y_{h}\}$ is harder to prove but some results can be obtained. Taking the region $\{y_{h}<y(s)\le2y_{h}\}$ as an example, we first note that for future directed causal curves crossing the horizon we must have $g(\xi_h,\dot{\gamma})<0$ implying again Eq.\eqref{cond1}, moreover Eq.\eqref{eqh2} is valid in the extremal case, so using the argument of \eqref{contra} we now find
\begin{equation}
\frac{d\hat{h}^2}{dy}\le-\frac{\varepsilon}{(y-y_h)^3},
\end{equation}
which by integration gives
\begin{equation}
\hat{h}^2(y_{i}^-)-\hat{h}^2(y_{i}^+)\le-\int_{y_{i}^+}^{y_{i}^-}\frac{\varepsilon}{(y-y_h)^3}dy<0,
\end{equation}
and thus $\dot{y}$ is negative close enough to the horizon in $\mathcal{M}_{I}$ and indeed positive in $\mathcal{M}_{IV}$ as claimed in the section above. Hence, by looking at the most singular term of Eq.\eqref{eqlo1}, then Eq.\eqref{eqh2} gives
\begin{equation}
\frac{d\hat{h}^2}{dy}=-\frac{2F(y_h)^3(RC(1+y_h)p_t-p_\psi)^2}{y_h^2(1-y_h^2)(y-y_h)^3}+\mathcal{O}((y-y_h)^{-2}),
\end{equation}
which by integration leads to
\begin{equation} 
\hat{h}^2=-\frac{F(y_h)^3(RC(1+y_h)p_t-p_\psi)^2}{y_h^2(y_h^2-1)(y-y_h)^2}+\mathcal{O}((y-y_h)^{-1}).
\end{equation}
From Eq.\eqref{eqh3} we now find
\begin{equation} \label{ybou3}
\dot{y}=\pm\frac{(x_h-y_h)^2(RC(1+y_h)p_t-p_\psi)}{R^2\sqrt{-F(y_h)}F(x_h)}+\mathcal{O}(|y-y_h|^{\frac{1}{2}}),
\end{equation}
where $x_h:=\lim_{y\to y_h}x(y)$, thus $|\dot{y}|$ is bounded away from zero. Note that we have written Eq.\eqref{ybou3} with a $\pm$ sign. In the present context we should take the $-$ sign, whereas in the region $\{\frac{y_{h}}{2}<y(s)<y_{h}\}$ we should take the $+$ sign. However, finding a uniform bound on $|\dot{\theta}|$ is more complicated. Since $|\dot{y}|$ is bounded away from zero, from \eqref{eqpm} multiplied by $(ds/dy)^2$ we find the bound \eqref{tz12}, for which integration now gives
\begin{equation} \label{explode}
|\theta(y)-\theta(y_h)|\le C_{16}ln|y-y_h|
\end{equation}
So, if maximality of the extension is to be shown, a more careful analysis is needed in order to constrain the bound \eqref{tz12} even further. However, assuming that $\dot{\theta}$ has a finite limit as $y\to y_h$, we can find uniform bounds on $|\dot{v}|,|\dot{\hat{\psi}}|$ by rewriting \eqref{eqpm} as \eqref{vbou1}, then as in the non-extremal case we obtain
\begin{equation} \label{limb3}
\lim_{y\to y_h}\alpha(x,y)=\pm \lim_{y \to y_h}\beta(x,y).
\end{equation}
Also, taking the $-$ sign above and using \eqref{vbou1}, we find that the limit
\begin{equation} \label{limb2}
\lim_{y\to y_h}\frac{\alpha(x,y)+\beta(x,y)}{G(y)}
\end{equation}
exists. Moreover since $\dot{y}<0$ and hence $\alpha(x,y)<0$, only the $-$ sign in \eqref{limb3} needs to be considered, otherwise the limit \eqref{limb2} would not exist. Thus, since the limit of \eqref{vbou1} does exist then by considering the expressions for $\dot{v}$ and $\dot{\hat{\psi}}$
\begin{equation} \label{exbou}
\begin{split}
\dot{v}=&-\frac{F(y)}{F(x)}p_t-\frac{RC(1+y)F(y)}{G(y)}\left[\beta(x,y)+\alpha(x,y)\right],\\
\dot{\hat{\psi}}=&\frac{F(y)}{G(y)}\left[\beta(x,y)+\alpha(x,y)\right],
\end{split}
\end{equation}
where $\alpha(x,y)$ and $\beta(x,y)$ are given by \eqref{albe}, we find bounds on both $|\dot{v}|$ and $|\dot{\hat{\psi}}|$. If we had instead considered outgoing $(u,\hat{\psi})$ coordinates we would find similar bounds. This then implies that geodesics are extendible at the surfaces $\{v=\pm\infty\}$ and $\{u=\pm\infty\}$. We could then follow these geodesics by systematically introducing ingoing and outgoing coordinates and proceeding as above. On the other hand, at the Killing horizon $y=y_h$, we have that $\hat{\psi}-\Omega_h v$ is constant and thus $\dot{\hat{\psi}}-\Omega_h \dot{v}=0$ at $y=y_h$. This implies that $p_t=0$ from Eq.\eqref{exbou}, thus choosing $v$ as a parameter along these geodesics we find that the null generator of the horizon is complete since there is no bifurcation point. In the non-extremal case we could have also covered the entire space-time with ingoing and outgoing coordinates and showed maximality in this way, as there would be a strict enough bound on $\dot{\theta}$.

Finally, to what concerns non-global hyperbolicity, it follows, as in the non-extremal case, due to the existence of white whole regions that separate each space-time region $\mathcal{M}_{n}$ for a specific value of $n$ from the remaining ones, that the space-time $(\tilde{\mathcal{M}},\tilde{g})$ is non-globally hyperbolic.

\section{Hidden symmetries of the dipole black ring} \label{hid}
If a $d$-dimensional space-time $(\mathcal{M},g)$ has less than $d-1$ commuting isometries and its associated Hamilton-Jacobi equation admits a separable solution, then the space-time under consideration may have a hidden symmetry related to a higher rank Killing tensor\footnote{For an overview on hidden symmetries see \cite{Kubiznak:2008qp} and for a short review relevant for what follows see \cite{Durkee:2008an}.}. In \cite{Durkee:2008an} it was shown, for singly and doubly spinning black rings, that the HJ equation admitted a separable solution only for zero energy null geodesics, which was associated with the existence of a conformal Killing (CK) tensor and a conformal Killing-Yano (CKY) tensor in a 4-dimensional space-time obtained by Kaluza-Klein reduction. The hint that lead to perform such reduction came from the fact that for a space-time with $d-2$ commuting Killing vector fields, which includes all known black ring solutions, the HJ equation if separable for null geodesics can be written as
\begin{equation} \label{kt}
K_{(1)}^{\mu\nu}(x)p_\mu p_\nu=K_{(2)}^{\mu\nu}(y)p_\mu p_\nu=c,
\end{equation}
for some constant $c$, where both $K_{1}$ and $K_{2}$ are CK tensors, satisfying the relation
\begin{equation} \label{kt2}
K_{(1)}^{\mu\nu}(x)-K_{(2)}^{\mu\nu}(y)=\hat{f}(x,y)g^{\mu\nu},
\end{equation}
for some function $\hat{f}(x,y)$. Then from a similar equation as Eq.\eqref{hj2}, one could check that the components $K^{tt}$ and $K^{ti}$ did not affect the value of $c$ since
\begin{equation} \label{ckc}
c=K^{\mu\nu}p_{\mu}p_\nu=K^{tt}E^2-2K^{ti}Ep_{i}+K^{ij}p_{i}p_{j}=K^{ij}p_{i}p_{j},
\end{equation}
where the indices $i,j$ range over $y,\psi,x,\phi$. This suggested that, even though the singly or doubly spinning black ring space-time did not admit a CK tensor, the space-time obtained by a Kaluza-Klein reduction along the time direction $\partial_t$ might, since the components $K^{tt}$ and $K^{ti}$ will be removed during such procedure.

Now, since we have found the same sort of separability in Sec.\ref{f0z} for the dipole black ring, we expect that the same kind of hidden symmetry to be present in this case. In fact we will show below that this is indeed the case.

\subsection{Kaluza-Klein reduction}
We want to dimensionally reduce the metric \eqref{ds1} along the $\partial_t$ direction using the usual Kaluza-Klein method. Note that $\partial_t$ is spacelike in the ergoregion so we are not performing a timelike reduction. Also, since the dipole black ring space-time is a solution to the action \eqref{act}, the dimensionally reduced 4-dimensional space-time will be a solution of Einstein-Maxwell-Dilaton theory plus an extra scalar field and an extra gauge field (see \cite{Ortin:gs}, section 11). Moreover, the dimensionally reduced metric will be related to the original metric in the same way as if we had only the Einstein-Hilbert term in the original action. Hence, we can take the same ansatz as for the neutral case \cite{Durkee:2008an}
\begin{equation} 
ds^2=e^{\Phi/\sqrt{3}}h_{ij}dx^i dx^j + e^{-2\Phi/\sqrt{3}}(dt+A_{i}dx^i)^2.
\end{equation}
Through comparison with the line element \eqref{ds1} we deduce that
\begin{equation}
e^{-2\Phi/\sqrt{3}}=\sqrt{\frac{F(x)}{-F(y)}},~A_{i}dx^{i}=RC(1+y)d\psi,
\end{equation}
and thus the dimensionally reduced metric is given by
\begin{equation} \label{ds4}
ds^{2}_{4}=h_{ij}dx^{i}dx^j=\Lambda^2(x,y)\left(\frac{dx^2}{G(x)}-\frac{dy^2}{G(y)}+\frac{G(x)}{F(x)^3}d\phi^2-\frac{G(y)}{F(y)^3}d\psi^2\right) ,
\end{equation}
where
\begin{equation}
\Lambda^2(x.y)=\sqrt{\frac{-F(y)}{F(x)}}\frac{R^2F(y)^2F(x)}{(x-y)^2}.
\end{equation}

Prior knowledge of this C-metric-like would lead to the dipole black ring metric simply by uplifting \eqref{ds4} to 5-dimensions by means of the inverse Kaluza-Klein procedure. This metric is in fact conformally related to a metric included in a general class of metrics found in \cite{Krtous:2008tb},\cite{Houri:2008ng}\footnote{I am grateful to Mark Durkee for pointing this out to me.}.

\subsection{Conformal Killing tensors and conformal Killing-Yano tensors}
First, note that $\Lambda^{2}(x,y)$ has the role of a conformal factor in the metric \eqref{ds4} since it is real and strictly positive everywhere inside the ergoregion $\{\frac{1}{\nu}<y<+\infty\}$. Therefore the zero energy null geodesics of the 5-dimensional dipole black ring correspond exactly to the null geodesics of this C-metric-like (compare $g^{\mu\nu}p_{\mu}p_{\nu}=0$ when $E=0$ with \eqref{ds4} in the case $ds_{4}^2=0$). Thus, we should expect a CK tensor in the lower dimensional metric by the analysis below Eq.\eqref{ckc}. In fact, we can read off from \eqref{hj2} the non-vanishing $K^{ij}_{(1)}$ and $K^{ij}_{(2)}$ components of the Killing tensors
\begin{equation}
\begin{split}
K_{(1)}^{xx}=&G(x),~K_{ (2)}^{yy}=G(y),\\
K_{(1)}^{\phi\phi}=&\frac{F(x)^3}{G(x)},~K_{(2)}^{\psi\psi}=\frac{F(y)^3}{G(y)},
\end{split}
\end{equation}
which with little effort, and noting that the metric \eqref{ds4} is diagonal, can be show to satisfy $K_{(1)}^{ij}-K_{(2)}^{ij}=\Lambda^2(x,y)h^{ij}$. Now, remembering Eq.\eqref{kt2}, $K_{(1)}$ and $K_{(2)}$ are not independent of each other, hence the most natural choice for a Killing tensor $K$ would be to take $K\equiv K_{(1)}+K_{(2)}$. It can then be shown that $K$ satisfies the conformal Killing equation
\begin{equation} \label{ck1}
\nabla_{(i}K_{jk)}=\hat{\omega}_{(i} h_{jk)},
\end{equation}
with 
\begin{equation}
\hat{\omega}=2\Lambda(x,y)\left(\frac{\partial \Lambda(x,y)}{\partial x}dx-\frac{\partial \Lambda(x,y)}{\partial y}dy\right),
\end{equation}
and therefore it is indeed a CK tensor. This calculation remains valid whatever conformal factor $\Lambda$ we could have taken, and in fact with indices raised $K=K^{ij}\partial_{i} \partial_{j}$ is diagonal and independent of the conformal factor.

We can now try and find a CKY tensor from  which $K^{ij}$ can be constructed from, that is, we look for a tensor $k^{ij}$ such that $K^{ij}=k^{ik}k^{jl}h_{kl}$. It is straightforward to see that this will be the case if we pick the tensor
\begin{equation}
k^{x\phi}=\frac{\sqrt{F(x)^3}}{\Lambda(x,y)}=-k^{\phi x},~k^{y\psi}=\frac{\sqrt{-F(y)^3}}{\Lambda(x,y)}=-k^{\psi y},
\end{equation}
with all other components vanishing. Lowering the indices we obtain the $2$-form
\begin{equation} \label{cky}
k=\Lambda^3(x,y)\left(\frac{1}{\sqrt{F(x)^3}}dx\wedge d\phi - \frac{1}{\sqrt{-F(y)^3}}dy\wedge d\psi\right).
\end{equation}
It can then be shown that $k$ satisfies the conformal Killing-Yano equation 
\begin{equation} \label{ck2}
\nabla_{(i}k_{j)k}=h_{ij}\hat{\xi}_k-\hat{\xi}_{(i}h_{j)k},
\end{equation}
with
\begin{equation}
\hat{\xi}=\frac{G(x)}{\sqrt{F(x)^3}}\frac{\partial \Lambda(x,y)}{\partial x}d\phi+\frac{G(y)}{\sqrt{-F(y)^3}}\frac{\partial \Lambda(x,y)}{\partial y}d\psi,
\end{equation}
and hence it is a conformal Killing-Yano tensor. As a final comment we note that taking the hodge dual of $k$ we can construct another CKY tensor which takes the same form as \eqref{cky}.

\section{Summary \& open questions} \label{far}
In this study we have shown that one can construct different types of extensions across the Killing horizons of non-extremal and extremal dipole black rings. In the non-extremal case we have show that the extension of Sec.\ref{snex} is maximal and timelike complete and furthermore, based on the results for the extremal case in Sec.\ref{sex}, that an alternative extension could be constructed by patching infinite copies of ingoing and outgoing coordinates. On the other hand, maximality in the extremal case remains an open and interesting problem, in fact, the kind of logarithmic behaviour of Eq.\eqref{explode} is normally encountered when constructing extensions in Taub-Nut space-times leading to the possibility of obtaining inequivalent extensions\footnote{I am grateful to Piotr T. Chru\'{s}ciel for pointing out this fact to me.}. This issue deserves further study. We have also successfully demonstrated that the causal structure in both cases resembles that of the 4-dimensional Reissner-Nordstr\"{o}m black hole, as expected, since the doubly and singly spinning black rings resemble that of 4-dimensional Kerr and Schwarzschild black holes respectively. This shows that, to what concerns the causal relations of the full space-time, there is no essential difference between these exotic ring solutions and the known 4d black hole solutions. Moreover, we have shown that the geodesic structure of the dipole black ring is very similar to that of the singly spinning black ring and that the dipole black ring also has the same type of hidden symmetry as the neutral case, which seems to be a general property of this type of geometries.  

Finally, it would be interesting to investigate how easily the techniques employed here can be used to show the maximality of the extension constructed in \cite{Chrusciel:2009vr} for doubly spinning black rings and also how all these considerations apply to different dipole black ring solutions of other gravity theories (for example, the ones found in \cite{Emparan:2006mm}, \cite{Yazadjiev:2005gs}) or more complicated solutions such as di-rings (\cite{Evslin:2007fv},\cite{Iguchi:2007is}) and bi-rings (\cite{Elvang:2007hs}).

\section*{Acknowledgements}
I am extremely grateful to Piotr T. Chru\'{s}ciel for nice discussions and very useful comments to an early draft of this paper, and to Mark Durkee for more useful comments. I also want to thank Julien Cortier for providing a technical note and to Micha\l$~$Eckstein and Alfonso García-Parrado Gómez-Lobo for letting me modify their original images.
I would also like to dedicate this work to a few people. To Harvey Reall and Hari Kunduri for the excellent lectures on black holes, to Troels Harmark for teaching me a great deal more on this topic and to Niels Obers for being an excellent supervisor and friend. To José Fonseca, João Laia and Francisco Gil Pedro for having been extremely patient with me during Part III and to Ricardo Monteiro, Jorge Santos and Miguel Paulos for having been so helpful during the same period. To Konstantinos Zoubos for being able to deal with my stupidity without laughing too much. To Conceição Nascimento for unquestionable support and to Pedro Lucas for being the most influential person (in a good sense) during the past 10 years. Without these people, I wouldn't had been able to write this paper. Finally, I would like to dedicate this paper to all those people who have fought for the preservation of Christiania. I am funded by FCT Portugal, even though I am from the Azores.

\bibliographystyle{utphys}
\bibliography{bibrings}

\end{document}